\documentclass[prd,twocolumn,tightenlines,superscriptaddress,nofootinbib,notitlepage,preprintnumbers]{revtex4-2}

        \usepackage[normalem]{ulem}       
      \usepackage{xcolor}

\usepackage[colorlinks=true, pdfstartview=FitV,linkcolor=blue, citecolor=blue, urlcolor=blue]{hyperref}

\usepackage{graphicx}
\usepackage[figuresleft]{rotating}

\usepackage[greek,british]{babel}

\usepackage{graphicx}
\usepackage{dcolumn}
\usepackage{bm}

\usepackage{latexsym,array,theorem,mathrsfs,subfigure,bm,float}
\usepackage{stackengine}
\usepackage{amsmath}
\usepackage{amssymb}
\usepackage{xcolor}
\usepackage{mathtools}
\usepackage{natbib}
\usepackage{tensor}
\usepackage[T1]{fontenc}
\usepackage{graphicx}

\newcommand{\be}{\begin{eqnarray}}
\newcommand{\ee}{\end{eqnarray}}

\newcommand{\F}{{B}}
\newcommand{\A}{{B}}

\newcommand{\WW}{{\rm W}}
\newcommand{\nn}{\nonumber}
\newcommand{\thetaw}{\theta_{\mbox{\tiny W}}}
\newcommand{\T}{{\rm T}}
\newcommand{\rr}{{\rm r}}

\newcommand{\U}{\mho}
\newcommand{\Uo}{{\rm U}}
\newcommand{\So}{{\rm S}}

\newcommand{\Ko}{{\rm K}}

\newcommand{\N}{{\rm N}}
\newcommand{\rrh}{{\rm r}_{\rm H}}

\newcommand{\xx}{\rr}

\newcommand{\X}{{x}}

\newcommand{\FF}{{\mathcal F}}
\newcommand{\J}{{\mathcal J}}

\newcommand{\mz}{m_{\mbox{\tiny Z}}}
\newcommand{\mh}{m_{\mbox{\tiny H}}}
\newcommand{\mw}{m_{\mbox{\tiny W}}}

\begin{document}

\title{Black Holes with Electroweak Hair
}

\author{Romain Gervalle}
\email{romain.gervalle@univ-tours.fr}
\affiliation{
Institut Denis Poisson, UMR - CNRS 7013, Universit\'{e} de Tours, Parc de Grandmont, 37200 Tours, France}

\author{Mikhail~S.~Volkov}
\email{mvolkov@univ-tours.fr}
\affiliation{
Institut Denis Poisson, UMR - CNRS 7013, Universit\'{e} de Tours, Parc de Grandmont, 37200 Tours, France}

\begin{abstract}
We construct static and axially symmetric magnetically charged hairy black holes in the gravity-coupled 
Weinberg-Salam theory.  Large black holes merge with the 
Reissner-Nordstr\"om (RN) family, while the small ones are extremal and 
support a hair in the form of 
a ring-shaped electroweak condensate carrying 
superconducting $W$ currents and up to $22\%$ of the total magnetic charge. 
The extremal solutions are asymptotically  RN 
with a  mass {\it below} the total charge, $M<|Q|$,  due to the negative Zeeman energy of the condensate 
interacting with the black hole magnetic field. Therefore, they cannot decay into RN black holes. As their charge increases, 
they show a phase transition when the horizon symmetry changes from spherical to oblate. At this point 
they  have the mass typical for  planetary size black holes of which $\approx 11\%$ 
is stored in the hair. 
Being obtained 
within a well-tested  theory, our solutions are expected to  be  physically relevant.

\end{abstract}

\maketitle

 In 1971 Ruffini and Wheeler formulated the no-hair conjecture according 
to which the only parameters of an isolated  black hole should be its mass, angular momentum, 
electric and/or magnetic charge \cite{Ruffini}. 
 The conjecture was supported by the uniqueness theorems in the electrovacuum sector 
(see \cite{Heusler1996} for a review) and by the no-hair theorems proven 
for a number of special cases (see, e.g., \cite{Bekenstein:1971hc,Saa:1996aw}). The  first 
broadly recognized 
counterexample to the conjecture was discovered in 1989 \cite{Volkov:1989fii}:  static 
black holes supporting   a nontrivial Yang-Mills field without a charge
 (see \cite{Volkov:2016ehx} for details).
 This finding was generalized 
in many ways, leading to a plethora of black holes supporting a Skyrme hair 
\cite{DROZ1991371}, a Higgs hair \cite{BREITENLOHNER1992357,Greene:1992fw}, 
stringy hair \cite{Kanti:1995vq}, hair without spherical symmetry \cite{PhysRevLett.79.1595}, and many 
others; see \cite{Volkov:1998cc} for a review. Some more recent examples include hairy black holes in the
Horndeski theory
\cite{PhysRevD.90.124063}, 
black holes supporting ``spinning clouds''  of  ultralight bosons \cite{Herdeiro:2014goa},
black holes with a massive graviton hair \cite{Brito:2013xaa}, etc.
\cite{Herdeiro:2015waa}. 

Nowadays  hairy black holes 
have become commonplace, 
but one may wonder 
if they really exist in nature ?
Unfortunately, all known solutions were obtained  within either simplified models or exotic models 
relying  on the existence of yet undiscovered particles and fields. 
Therefore, their physical relevance is not immediately obvious. 
To be relevant, a solution
should be obtained within a well-tested  physical theory, which is 
the standard model (SM) of fundamental interactions coupled to the 
general relativity (GR). The former contains the QCD sector where 
quantum effects are dominant, but in 
the electroweak (EW) sector the quantum corrections are not too large. 
Therefore, it makes sense
to study extended  classical configurations  in  the 
Einstein-Weinberg-Salam (EWS) theory. 
This theory contains the electrovacuum  sector with the Kerr-Newman black holes,
but  does it allow for some other black holes as well ?

In fact, such solutions do exist but are difficult to construct
since magnetically charged hairy EW black holes are not spherically 
symmetric, unless in an unphysical limit of the theory \cite{Greene:1992fw}, or for the low 
value of the magnetic charge  \cite{Bai:2020ezy}. 
Nonspherical  black holes were considered  perturbatively in theories 
which are similar to the EWS but without  the $Z$  boson \cite{Lee:1994sk},
or  without both $Z$ and Higgs bosons  \cite{Ridgway:1995ke}.
A fully nonperturbative analysis within the EWS theory has been missing up to now.  

Maldacena has qualitatively described such black holes  \cite{Maldacena:2020skw}. They 
are not spherically symmetric because their strong magnetic field  creates 
the EW condensate of vortices  \cite{Ambjorn:1988tm,Ambjorn:1989sz,Chernodub:2012fi,Chernodub:2022ywg} 
 forming a ``corona'' around the black hole. This
happens in the region where the hypermagnetic field ${\rm B}$  falls within the interval $\mh^2>{\rm B}>\mw^2$.
If ${\rm B}>\mh^2$ at the horizon, then the black hole is surrounded in addition by a bubble of false vacuum.
 In the far field all massive fields assume vacuum values and the magnetic field 
 becomes spherically symmetric. 
 
 Following this scenario, various aspects of magnetic black holes have been discussed, but always at the qualitative level 
 \cite{Bai:2020spd,Bai:2021ewf,Ghosh:2020tdu,Estes:2022buj,Diamond:2021scl}. Therefore, 
in what follows we confirm the scenario by explicitly constructing the solutions in the special  case of axial symmetry when the 
EW condensate reduces to rings in the equatorial plane. Analyzing the  general case 
would require 
3D simulations. 
We  sketch below only  essential points and systematically refer to the Supplemental Material  \cite{S} for details; 
an extended version will be presented separately \cite{GVin}. 

{\bf The theory.} 
We consider the bosonic part of the EWS theory containing 
gravity described by the metric ${\rm g}_{\mu\nu}$, a complex doublet Higgs field  $\Phi$, 
a U(1) hypercharge field $\A= \A_\mu dx^\mu$,
and an SU(2) field $\WW=\T_a\WW^a_\mu dx^\mu$ with 
$\T_a=\tau_a/2$ where $\tau_a$ are Pauli matrices. 
The action is 
\be                                     \label{I}
{\mathcal S}=\frac{e^2}{4\pi\alpha}\int \left(\frac{1}{2\kappa}\,R-\frac{1}{4g^2}\,\WW^a_{\mu\nu}\WW^{a\mu\nu}
-\frac{1}{4g^{\prime 2}}\,{\F}_{\mu\nu}{\F}^{\mu\nu}\right. \nn \\
\left.-(D_\mu\Phi)^\dagger D^\mu\Phi
-\frac{\beta}{8}\left(\Phi^\dagger\Phi-1\right)^2\right)\sqrt{-\rm g}\, d^4x.~~~~
\ee
Here $R$ is the Ricci scalar, ${\F}_{\mu\nu}=\partial_\mu{\A}_\nu
-\partial_\nu{\A}_\mu$ and 
$\WW^a_{\mu\nu}=\partial_\mu\WW^a_\nu
-\partial_\nu \WW^a_\mu
+\epsilon_{abc}\WW^b_\mu\WW^c_\nu$
are the U(1) and SU(2) gauge field strengths, the Higgs 
covariant derivative is 
$
D_\mu\Phi
=\left(\partial_\mu-\frac{i}{2}\,{\A}_\mu
-\frac{i}{2}\,\tau_a \WW^a_\mu\right)\Phi,
$
and the Higgs self-coupling is $\beta\approx 1.88$. 
The two coupling constants 
$g=\cos\thetaw$ and
$g^\prime=\sin\thetaw\approx \sqrt{0.22}$ determine  the electron charge   $e=gg^\prime$ and 
$\alpha\approx 1/137$ is the fine structure constant. 
Everything is dimensionless, the $Z,W,$ and Higgs 
masses are $\mz=1/\sqrt{2}$, $\mw=g\mz$, $\mh=\sqrt{\beta}\mz$ in units of 
the mass scale $128.9 {\rm ~GeV}$,
whose  Compton wavelength 
$1.53\times 10^{-16}$cm  sets the length  scale. 
The gravity  coupling is $\kappa=(4e^2/\alpha) ({\mz}/{M}_{\rm Pl})^2=5.30\times 10^{-33}$ where 
${M}_{\rm Pl}$ is  the Planck  mass \cite{S}.

The  electromagnetic field 
is 
$e \FF_{\mu\nu}=g^2 \A_{\mu\nu}-g^{\prime 2}n_a\WW^a_{\mu\nu}$ with $n_a=(\Phi^\dagger\tau_a\Phi)/(\Phi^\dagger\Phi)$ and 
the electric  current is 
 $4\pi \J^\mu =\nabla_\nu \FF^{\mu\nu} $ \cite{Nambu:1977ag}. 
 Assuming the spacetime to be static and choosing the coordinates $x^\mu=(x^0,x^k)$ such that the timelike 
 Killing vector has components $\delta_0^\mu$, 
the dual tensor $\tilde{\FF}^{\mu\nu}$ defines the magnetic field ${\cal B}^k=\tilde{\FF}^{0k}$ and 
 the magnetic charge density $4\pi \tilde{\J}^0 =\nabla_k {\cal B}^k$. 
 The hypermagnetic field is ${\rm B}^k=\tilde{B}^{0k}$. The closed part of $\FF_{\mu\nu}$ 
 defines similarly a current 
  $J^\mu$ \cite{S}. 
 
 {\bf Abelian  solutions.}
 Varying the action gives the field equations \cite{S}, whose simplest solution corresponds to 
 a vacuum geometry with $\A=\WW=0$ and $\Phi^{\rm tr}=(0,1)$. 
 Choosing  the gauge fields within  the Cartan subalgebra, gives the {\it Abelian} solution with the RN geometry, 
 \be                 \label{RN} 
 ds^2=-N(r)\, dt^2+\frac{dr^2}{N(r)}+r^2(d\vartheta^2+\sin^2\vartheta\,d\varphi^2),~~~~~ \\
 \A=-(n/2)\,\cos\vartheta\, d\varphi,
 ~~\WW=\T_3 \A,~~~\Phi^{\rm tr}=(0,1), \nn
 \ee
 with $N(r)=1-2M/r+Q^2/r^2$ 
 and $Q^2=(\kappa/2)\,P^2$ 
 where $P=n/(2e)$ and $n\in\mathbb{Z}$. 
 The gauge fields show the Dirac string 
 singularity at $\vartheta=0,\pi$, which can be removed by 
 gauge transformations. The radial hypermagnetic field is ${\rm B}=n/(2r^2)$ while the  magnetic 
 field ${\cal B}=P/r^2={\rm B}/e$  corresponds to the 
 Dirac monopole.
 The solution \eqref{RN} describes the magnetic  RN black hole of charge $P$. 
 The event horizon is at $r_h=M+\sqrt{M^2-Q^2}$. 
 
 Another simple solution is obtained by setting in \eqref{RN} $\WW=\Phi=0$
 while keeping the same $\A$ and choosing  
 $N(r)=1-2M/r+g^2Q^2/r^2-(\Lambda/3)\, r^2 $ with $\Lambda=\kappa\beta/8$. 
 This time the magnetic field is ${\cal B}=g^2P/r^2=g^2{\rm B}/e$; hence, the solution 
 describes the RN-de Sitter  (RNdS) black hole of charge $g^2P$. 
 In the extremal limit the two black hole horizons merge, then 
 \be           \label{RNdS}
 N(r)=\left(1-\frac{r_{\rm ex}}{r}\right)^2\left(1-\frac{\Lambda}{3}\left[r^2+2rr_{\rm ex}+3r_{\rm ex}^2\right]\right),
\ee
where $1-2\Lambda r_{\rm ex}^2=\sqrt{1-Q^2/Q_{\rm m}^2}$ with $Q_{\rm m}=1/(2g\sqrt{\Lambda})$. 
\begin{figure}[!b]
    \centering
    
      
       \includegraphics[scale=0.32]{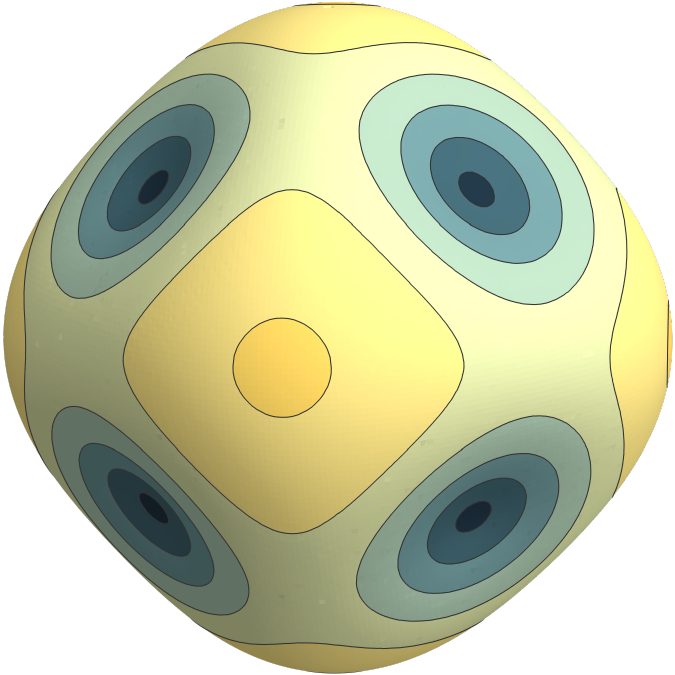}
      \hspace{2 mm}
      \includegraphics[scale=0.32]{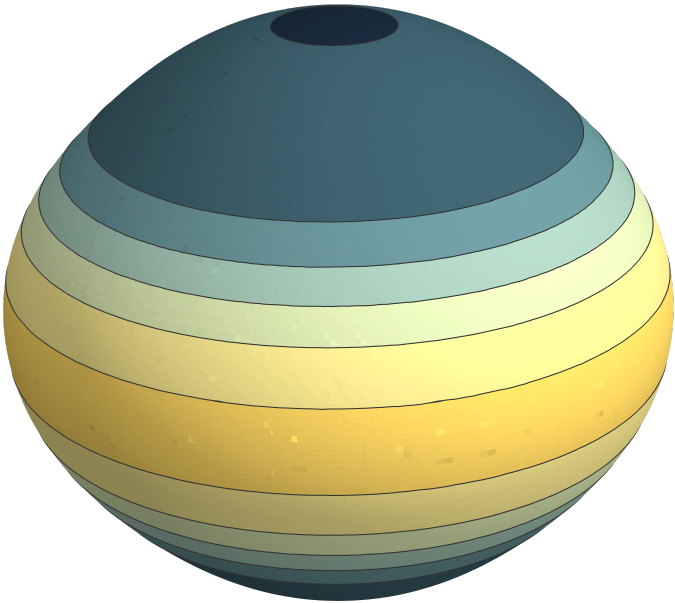}

       \caption{\small        Left:  the horizon distribution of the $W$ condensate 
$\bar{w}^\mu w_\mu$  minimizing the energy for $n=10$. The level lines coincide with the current 
flow forming loops around radial vortex pieces (dark spots). 
The condensate and vortex fields are maximal in the yellow regions. 
Right: the same when 
all vortices merge into two oppositely directed multivortices. 
        }
    \label{Fig0}
\end{figure}

 {\bf Perturbative analysis.} 
 Perturbing all fields  around the RN solution \eqref{RN} and 
linearizing with respect to perturbations, the field
$w_\mu=(\delta {\rm W}^1_\mu+i\,\delta {\rm W}^2_\mu)/g$ decouples from the rest (in the unitary gauge)
and fulfills the equations which admit  the following solution \cite{S}:
\be                      \label{w} 
w_\mu dx^\mu=  e^{i\omega t}\psi(r)\, 
(\sin\vartheta)^{j+1}\sum_{{\rm m}\in[-j,j]} c_{\rm m}z^{\rm m-1} dz.~~~~
\ee
Here
 $j=|n|/2-1$ is the orbital angular momentum,
 $|n|>1$ (if  $|n|=1$ then $w_\mu$  is unbounded at $\vartheta=0,\pi$), 
$z=\tan(\vartheta/2) \exp(-in\varphi/|n|)$, while  $c_{\rm m}$ are arbitrary 
 coefficients. The function $\psi(r)$ fulfills the equation 
\be             \label{RNs}
\left(-\frac{d^2}{dr_\star^2}+N(r)\left[\mw^2-\frac{|n|}{2r^2}\right]\right)\psi(r)= \omega^2\psi(r),
\ee
with $dr_\star=dr/N(r)$.  If $N(r)=1$,  this equation always admits 
normalizable solutions with $\omega^2<0$; hence, Dirac monopoles in flat space are always unstable 
with respect to the formation of a $W$ condensate \cite{GVI}.
If $N(r)$ corresponds to the RN solution,  the instability is absent for a large  horizon radius $r_h$, 
but settles in for $r_{h}=r^0_h(n)$ when the equation admits a solution with $\omega=0$
describing a static deformation
of the black hole by a condensate. Therefore,  $r_h^0$  is  the 
bifurcation point where the first hairy black hole solutions deviate from the RN family.

Solving Eq.\eqref{RNs} numerically yields 
 $r_h^0(n)=$
$0.89$, $2.68$, $10.29$  for $n=2,10,100$, respectively. 
For $|n|\gg 1$ one has 
$r_h^0(n)\approx 1.13\,\sqrt{|n|}=\sqrt{|n| }/g$; hence, 
${\rm B}(r^0_h)\approx \mw^2$. 
If $|n|=2$ then $j=0$ and 
the hair  is spherically symmetric, but if $|n|>2$ then  $j>0$ and  the hair is 
not  spherical.

The coefficients $c_{\rm m}$ are fixed by going to higher orders of perturbation theory and minimizing the condensate energy \cite{S}. 
This gives the result shown in  Fig.\ref{Fig0} (left): a ``platonic'' distribution  of radial vortex pieces forming  the corona. 
The vortices extend along the radial direction and 
 are encircled by loops of  current  $4\pi J_\nu=e\Im \left[\nabla^\sigma(\bar{w}_\sigma w_\nu)\right]$  tangent 
to the horizon.
According to the EW anti-Lenz  law  \cite{Ambjorn:1988tm,Ambjorn:1989sz,Chernodub:2012fi,Chernodub:2022ywg}, 
 the vortex fields  vanish at the vortex center where the condensate vanishes and become  maximal 
where the condensate is maximal.

For large $|n|$ the condensate has at most only discrete symmetries, although the background 
magnetic field creating it 
is spherically symmetric. It is interesting that a similar phenomenon of symmetry breaking was observed  in  liquid helium, 
where the minimal energy defects created by applying an external field do not share the same symmetry with the field applied \cite{Thuneberg,Salomaa,Salomaa:1987zz}.

The choice $c_{\rm m}=\delta_{\rm m 0}$ is also a critical point of the energy, although not the global minimum. 
It corresponds to the axially symmetric configuration in which 
all vortices merge into two oppositely directed multivortices supported by
azimuthal currents; see Fig.\ref{Fig0} (right).

Summarizing,  the perturbative analysis shows zero modes describing the hair that starts to grow. 
The hairy solutions  are 
not spherically symmetric for $|n|>2$, but they can be axially symmetric, and these we are able to 
construct also  at the nonperturbative level.

{\bf Axial symmetry.} We choose the fields as follows:
\be                   \label{metr}
ds^2&=&-e^{2\Uo} {\rm N}(\rr)\, dt^2+e^{-2\Uo} dl^2, \\
dl^2&=&
e^{2\Ko}\left[\frac{d\rr^2}{{\rm N}(\rr)}+\rr^2d\vartheta^2\right]+e^{2 \So}\,\rr^2\sin^2\vartheta d\varphi^2,\nn \\
\WW&=&\T_2\left( F_1\,d\rr+F_2\,d\vartheta \right)-\frac{n}{2}\left( \T_3\,F_3-\T_1 F_4\,\right)d\varphi\,, \nn \\
\A&=&-(n/2)\, Y\,d\varphi\,,~~~\Phi^{\rm tr}=(\phi_1,\phi_2). \nn 
\ee
Here $\Uo,\Ko,\So,F_1,F_2,F_3,F_4,Y,\phi_1,\phi_2$ are 10 real functions of $\rr,\vartheta$
subject to certain boundary conditions \cite{S}. 
The gauge fields $\WW,\A$  show the Dirac string singularity that  can be gauged away 
if $n\in\mathbb{Z}$ \cite{S}. They also have a residual U(1) gauge invariance that can be fixed by imposing a gauge condition \cite{S}. 
The radial coordinate $\rr$ (not the same as the Schwarzschild  coordinate $r$) is 
defined up to reparametrizations, its choice is specified by fixing  the auxiliary function $\N(\rr)$. 
For nonextremal solutions we set $\N(\rr)=1-\rrh/\rr$ where  the black hole ``size'' $\rrh$ does not have a direct physical meaning 
but labels the solutions. 
We obtain solutions for the 10 functions in \eqref{metr} for values of the gravity coupling $10^{-10}<\kappa<10^{-2}$ and then extrapolate 
to the physical value $\kappa\sim 10^{-33}$. 
We use the 
FreeFem  numerical solver \cite{MR3043640}.

\begin{figure}[!b]
    \centering
       \includegraphics[scale=0.5]{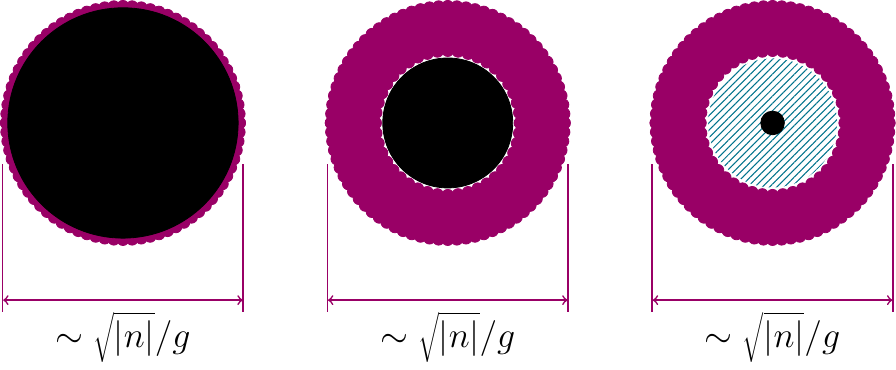}
 
       \caption{\small 
       When  the horizon size decreases, 
 the massive hair emerges (left), grows longer (center), then 
a bubble of symmetric phase appears around the horizon (right). 
        }
    \label{Fig22}
\end{figure}

{\bf Hairy black holes.} 
These black holes have the same charge as in the RN case, $P=n/(2e)$.
Choosing
a value $|n|>2$ (such that $Q=\sqrt{\kappa/2}\,P\ll Q_{\rm m}$) determines 
the size $\rrh=\rrh^0(n)$ 
for which the RN solution starts to develop a hair \cite{S}. 
Next, we decrease   the size  $\rrh$ and the 
horizon shrinks, but the massive hair grows  longer spanning 
 the interval $\rr\in[\rrh,\rrh^0]$. 
When $\rrh$ descends below a 
certain value $\rrh^{\rm b}$ such that ${\rm B}(\rrh^{\rm b})\approx \mh^2$,  
the massive hair does not grow anymore and  rests confined in the region 
$\rrh^{\rm b}<\rr<\rrh^0$ where $\mh^2>{\rm B}>\mw^2$. 
In the near-horizon region, 
$\rrh<\rr<\rrh^{\rm b}$, the hypermagnetic field becomes too strong, ${\rm B}>\mh^2$, 
driving to zero $\WW$ and $\Phi$ and creating 
a bubble of symmetric phase around the horizon (see Fig.\ref{Fig22}). 
The bubble expands as $\rrh$  decreases further, and in the $\rrh\to 0$ limit 
the horizon surface gravity approaches zero but the horizon area  remains  finite.

\begin{figure}[!b]
    \centering
           \includegraphics[scale=0.55]{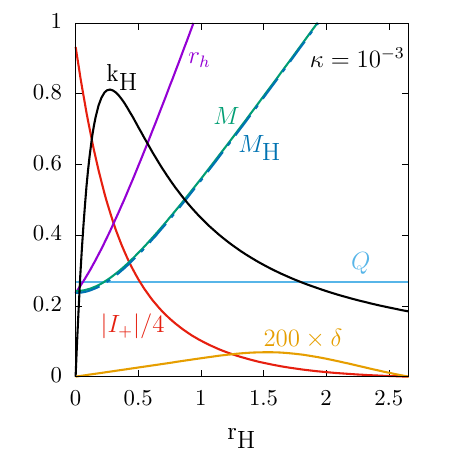}
    \includegraphics[scale=0.55]{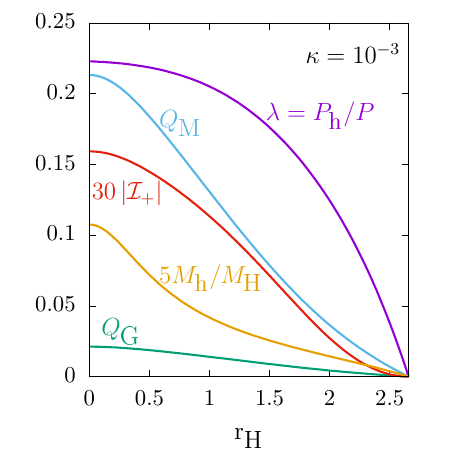}

       \caption{
       Parameters of nonextremal solutions with $n=10$, $\kappa=10^{-3}$. 
       For $\rrh\to 0$ they become extremal, for $\rrh\to 2.66$ they loose hair 
      and become RN.  
         }
    \label{Fig22a}
\end{figure}

Using the magnetic current density, one can compute 
the charge contained in the  hair, 
$P_{\rm h}=\int_{\rr>\rrh} \tilde{\mathcal{J}}^0\sqrt{-\rm g}\, d^3x$, while the rest $P_{\rm H}=P-P_{\rm h}$ is 
inside the black hole. Defining $\lambda=P_{\rm h}/P$, we observe that $\lambda\to 0$ 
at the bifurcation point 
 while in the extremal limit one has 
$\lambda\to g^{\prime 2}$; hence, $P_{\rm h}=g^{\prime 2}P=0.22\times P$ (see Fig.\ref{Fig22a}). 

Summarizing, hairy black holes exist for $\rrh\in [0,\rrh^0]$. They loose hair in the RN limit $\rrh\to \rrh^0$  
when all the charge is inside the black hole, 
and they become  maximally hairy in the extremal limit $\rrh\to 0$ when 
$22\%$ of the magnetic charge moves to  the hair.

The mass $M$ is determined from the asymptotic  expansion ${\rm g}_{00}=-1+2M/\rr+{\cal O}(1/\rr^2)$
or from the  formula 
\be          \label{mass} 
M=\frac{{\rm k}_{\rm H}{\rm A}_{\rm H}}{4\pi}+\frac{\kappa}{8\pi}\int_{\rr>\rrh}\left(-T^0_{~0}+T^k_{~k}\right)\,\sqrt{-\rm g}\, d^3x. 
\ee
Here 
${\rm k}_{\rm H}=(1/2)\left.\N^\prime e^{2\Uo-\Ko}\right|_{\rr=\rrh}$ 
is the surface gravity and ${\rm A}_{\rm H}=\left.2\pi \rrh^2\int_0^\pi e^{\Ko+\So-2\Uo }\sin\vartheta d\vartheta\right|_{\rr=\rrh}$ is the 
horizon area, and $T^\mu_{~\nu}$ is the stress-energy  tensor.
The total mass splits as $M=M_{\rm H}+M_{\rm h}$ where the horizon contribution $M_{\rm H}$ is defined as the mass of the RN 
black hole with the same area ${\rm A}_{\rm H}$ and with the  charge $P_{\rm H}$ \cite{S}. The mass $M$ decreases  with decreasing 
$\rrh$ and, unless $|n|$ is very large, one always has $M\approx M_{\rm H}$. 
The relative contribution of the hair mass
$M_{\rm h}/M_{\rm H}$  grows with $n$, it is maximal in the extremal limit and vanishes in the RN limit (see Fig.\ref{Fig22a}).

We define the  horizon radius as  $r_{h}=\sqrt{{\rm A}_{\rm H}/(4\pi)}$, but since the horizon is nonspherical, one should 
define also the equatorial radius 
$r_{h}^{\rm eq}=\sqrt{{\rm g}_{\varphi\varphi}(\rrh,\pi/2)}$ 
and the polar radius 
$r_{h}^{\rm pl}=(1/\pi) \int_0^\pi \sqrt{{\rm g}_{\vartheta\vartheta}(\rrh,\vartheta)}\,d\vartheta$,
 which determine the horizon 
oblateness $\delta =r_{h}^{\rm eq}/r_{h}^{\rm pl}-1$.

Far away from the horizon the massive fields decay and the theory reduces to electrovacuum;
hence, one can define \cite{S} the gravitational $Q_{\rm G}$ and magnetic $Q_{\rm M}$  quadrupole moments
following the approach of \cite{Geroch:1970cd,Hansen,Fodor,Fodor1}. 
As seen in Fig.\ref{Fig22a}, the quadrupole moments  grow 
as $\rrh$ decreases, reaching maximal values in the extremal  limit. 
The horizon oblateness $\delta$ also grows at first but then 
decreases and  vanishes in the extremal limit, when the horizon geometry  becomes   perfectly spherical,
although the hair is oblate. 

{\bf Extremal solutions.}  Extremal black holes have zero surface gravity.
The value of their charge 
$Q_\star\approx 0.6\, Q_{\rm m}$ with $Q_{\rm m}=1/(2g\sqrt{\Lambda})$ separates two phases, 
with the horizon oblateness $\delta$  being the order parameter. 

Phase I: $|Q|<Q_\star$, $\delta=0$.  The solutions interpolate between 
the extremal RNdS described by Eq.\eqref{RNdS} at the horizon and the RN given by Eq.\eqref{RN} in the asymptotic region. 
We obtain them by choosing  the auxiliary metric function 
$\N(\rr)\propto (1-r_{\rm ex}/\rr)^2$  \cite{S}. 
The horizon radius is $r_{\rm ex}=g|Q|+{\cal O}(|n|^3\kappa^{5/2})$ and the area ${\rm A}_{\rm H}=4\pi r_{\rm ex}^2$. 
\begin{figure}
    \centering
				\includegraphics[width=8 cm,angle =0 ]{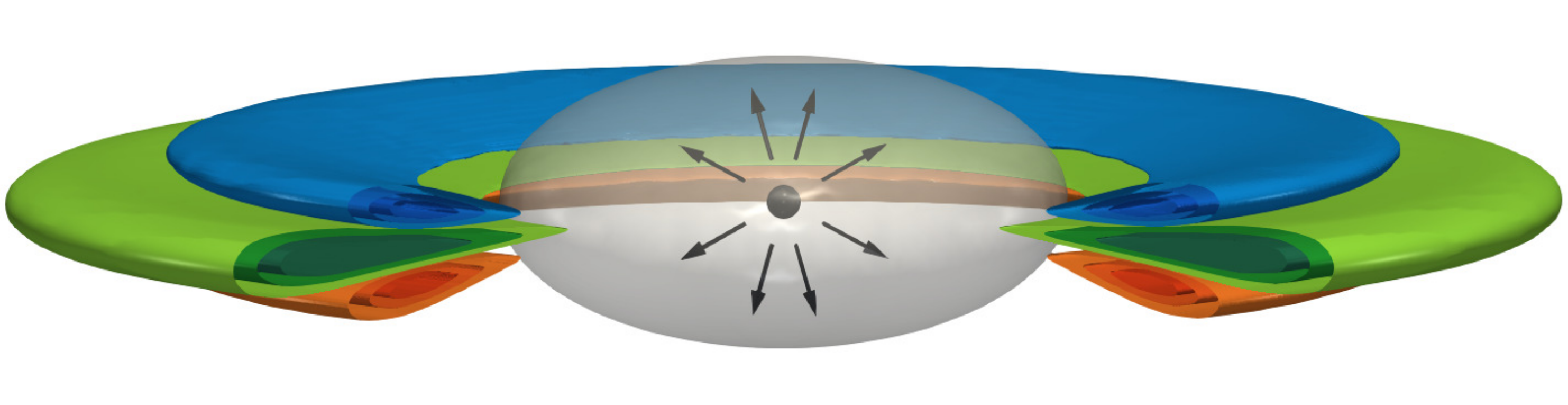}
						
       \caption{
The extremal  solutions contain  a small charged black hole inside a bubble of symmetric phase, surrounded 
by a ring-shaped EW condensate supporting $22\%$ of the total magnetic charge and two opposite superconducting $W$ currents. 
These create pieces of two magnetic multivortices along the positive and negative $z$ directions. Farther away the condensate 
disappears and 
the magnetic field becomes radial. 
        }		
    \label{ex}
\end{figure}

The configurations  harbor  at the center a small  black hole containing $78\%$ of the total magnetic charge and 
supporting a strong ${\rm B}$ field which creates  
a bubble of symmetric phase where $\WW\approx\Phi\approx 0$. The bubble spans the region 
of  size $r_{\rm b}\approx \sqrt{|n|/\beta}\approx 0.73\sqrt{|n}|$ where ${\rm B}> \mh^2$. 
 Farther away, where $\mh^2>{\rm B}> \mw^2$, the massive fields 
deviate from zero and form a ring-shaped condensate, as shown in  Fig.\ref{ex}. The central ring  (green online) 
contains  the remaining $22\%$ of the charge whose distribution is described by 
the magnetic charge density $\tilde{\mathcal{J}}^0$. This ring is sandwiched between two others (blue and red online) 
containing  equal but oppositely directed superconducting currents 
$\mathcal{I}_{+}$ and $\mathcal{I}_{-}$ (in units of $1.4\times 10^8$A) which are 
fluxes of $\J^\varphi$ in the upper and lower half-spaces. 
Fluxes of $J^\varphi$, $I_\pm$,  look similar  but have a much larger amplitude (a more detailed description will be given in \cite{GVin}).

The condensate 
terminates where ${\rm B}\approx\mw^2$  at $r_{\rm c}\approx \sqrt{|n|}/g\approx 1.13\sqrt{|n|}$, which determines
the corona size. 
Still farther away 
the configuration approaches the RN solution \eqref{RN} whose  mass is 
{\it less} than the charge,
\be
M<|Q|.
\ee
This is because, although  the hair carries the charge
 $Q_{\rm h}=0.22\times Q$, its mass $M_{\rm h}$ is small  due to the negative Zeeman energy of the 
condensate  interacting with the magnetic field of the black hole, which shifts the $W$ mass as $\mw^2\to \mw^2-|{\rm B}|$ \cite{Ambjorn:1989sz}. 
As a result, the 
mass-to-charge  ratio for the hair is small, 
$M_{\rm h}/|Q_{\rm h}|\sim \sqrt{\kappa}\ll 1$, which can be viewed as a manifestation 
of the weak gravity conjecture \cite{Arkani-Hamed:2006emk}. The condensate is magnetically repelled by the black hole
stronger than attracted gravitationally,  but it cannot fly away because it has to obey  the Yukawa law. 
Since the hair mass is small, one has 
$M=M_{\rm H}+M_{\rm h}\approx M_{\rm H}=(r_{\rm ex}+g^2Q^2/r_{\rm ex})/2\approx g|Q|<|Q|$. 

\begin{figure}[!b]
    \centering
    
     \includegraphics[scale=0.549]{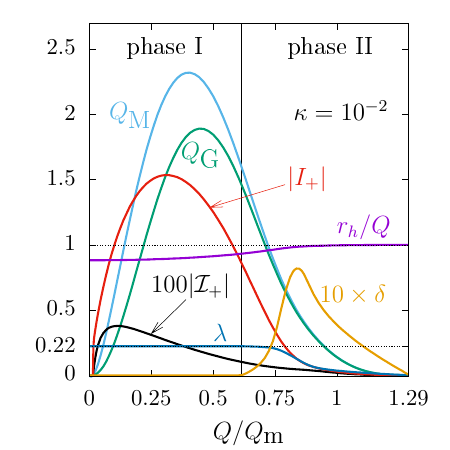}
     \includegraphics[scale=0.549]{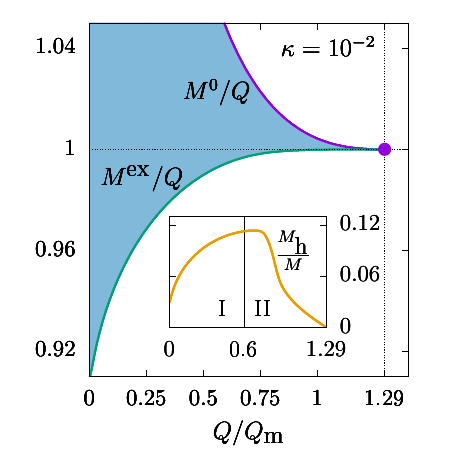}
               
       \caption{
 The parameters of  extremal  solutions (left) and the ratio $M/|Q|$  for the hairy solutions (right)
 for $\kappa=10^{-2}$. The insertion in the right panel shows the fraction of the mass stored in the hair for the extremal solutions.
        }
    \label{Fig23}
\end{figure}

 Extrapolating our results 
  to the physical value 
 $\kappa\sim 10^{-33}$   yields the following picture. The horizon size $r_{\rm ex}\propto Q\propto \sqrt{\kappa}$ 
 is {\it parametrically} small as compared to the size of the hairy region $\sim \sqrt{|n|}$.  
 Therefore,  the tiny black hole does not affect the electroweak hair  but only creates  the ${\rm B}$ field 
 which produces the condensate and accounts
 for the horizon part of the mass, $M_{\rm H}=g|Q|\propto \sqrt{\kappa}$. 
 The condensate  carries the mass $M_{\rm h}\sim \kappa$ 
 and lives far away from the horizon, in the region where 
 the geometry is almost flat. 
 The total mass in units of the electroweak mass scale 
 is $e^2/(4\pi\alpha)\times{\cal M}$, where \cite{S} 
  \be
 {\cal M}\equiv \frac{8\pi}{\kappa}\times  M=\frac{4\pi g|n|}{\sqrt{2\kappa}\,e}+\frac{8\pi}{\kappa}\,M_{\rm h}\equiv {\cal M}_{\rm H}+{\cal M}_{\rm h}. 
 \ee
 The  horizon contribution  ${\cal M}_{\rm H}$ corresponds to the mass $5.1\,|n|\,M_{\rm Pl}$, which 
  diverges when  $\kappa\to 0$, but  the hair contribution 
 ${\cal M}_{\rm h}$ remains then finite  and reduces precisely to the regularized mass of the flat space EW monopoles:
 the Cho-Maison monopole \cite{Cho:1996qd} for $|n|=2$ and its  generalizations for $|n|>2$ \cite{GVII}. 
 As a result, extremal solutions can be viewed
 as flat space monopoles harboring in the center a tiny RNdS black hole.
 The total mass of the  flat space EW  monopoles
 is infinite; hence, it costs infinite energy to create them, 
 but gravity renders the mass  finite due to the cutoff at the horizon. Therefore, the extremal black holes 
 provide the ``classical UV completion'' for the EW monopoles.

When increasing  the charge $Q\propto |n|$, the bubble size and hair length scale as $\sqrt{|Q|}$,
but the horizon size  $r_{\rm ex}\propto |Q|$ grows faster and the black hole absorbs the bubble. 
The 
hair mass $M_{\rm h}$ 
grows faster than the horizon mass 
$M_{\rm H}$ 
increasing 
the ratio $M/|Q|$. 
The horizon value of the ${\rm B}$ field is $\propto Q/r_{\rm ex}^2\propto 1/Q$; hence, it {\it decreases}, 
and when ${\rm B}<\mh^2$, the Higgs field deviates from zero at the horizon, the bubble is totally absorbed
and the solution enters next phase. \\
Phase II: $|Q|>Q_\star$, $\delta>0$. 
The horizon geometry changes from spherical to oblate and is no longer described by the extremal RNdS solution \eqref{RNdS} \cite{S}. 
One has 
$\delta\propto (|Q|-Q_\star)^s$ at the transition point, which resembles 
a second order phase transition (the critical exponent is rather large, $s\approx 10.8$ if $\kappa=10^{-2}$). 
In addition, the fraction of the hair charge which was constant in the previous phase, $\lambda=P_{\rm h}/P=0.22$, 
starts to decrease (see Fig.\ref{Fig23}). 
The black hole absorbs  the corona and becomes less hairy, the ratio $M/|Q|$ increases,
the geometry approaches the extremal RN
and finally merges with it when the horizon size overtakes the corona size, then 
$|Q|=r^0_h(n)=\sqrt{|n|}/g$. This corresponds to the maximal charge 
 $Q_{\rm max}=2g^\prime\sqrt{\beta}\,Q_{\rm m}=1.29\,Q_{\rm m}$. No hairy solutions exist beyond this value of charge.

 The black hole is maximally hairy for $Q\approx Q_\star$ ($|n|\approx 1.5 \times 10^{32}$) 
 when the  ratio $M_{\rm h}/M$ 
 is maximal (see Fig.\ref{Fig23}).   The horizon size is then $\approx 1.37~{\rm cm}$ 
and  the  total mass $\approx 2\times 10^{25}~{\rm kg}$ is typical for the planetary mass black holes,  of which $\approx 11\%$ (if $\kappa=10^{-2}$) 
is stored in the hair.

{\bf Summary.} We obtain the hairy solutions  within the region 
corresponding to the shaded area  in 
 Fig.\ref{Fig23} (right panel):   below the upper curve corresponding to
 the bifurcation with the RN branch and above the lower curve corresponding to the extremal solutions. 
 Solutions for $|n|>2$ are new. 
  The corona greatly enhances the evaporation rate \cite{Maldacena:2020skw};
 hence, nonextremal holes should quickly relax to the extremal state where $M\leq |Q|$, therefore they cannot decay into RN black holes.
 
 However, they can still lower their energy by developing a ``hedgehog'' of radial vortices   forming  the corona. 
 In the axially symmetric case the corona degenerates  into two oppositely  directed multivortices which are 
 unstable due to the repulsion between  the elementary  vortices. 
 The axially symmetric corona can lower its energy via splitting 
 into elementary vortices with maximal mutual separations, 
 similarly to what is shown in Fig.\ref{Fig0}.  
 The perturbative analysis indicates that such a ``spread'' configuration of vortices  
 achieves the absolute energy minimum \cite{S}. 
 Therefore, the corresponding ``platonic''  black holes are expected to be stable, 
 although nonaxially symmetric solutions describing such black holes are not yet known
 and their systematic stability analysis is still to be carried out.

Since they are described by well-tested theories, 
the hairy EW  black holes are expected to be physically relevant. 
They  could probably originate from  primordial black holes. 
Suppose that the fluctuating hypermagnetic field  in the  ambiant  EW plasma  becomes 
 at some moment  mostly orthogonal to the black hole horizon, or that a piece of a magnetic vortex gets attached to the horizon. 
 This will create a flux through the horizon, hence a charge, to be compensated by the 
opposite flux  created on the other black hole(s). 
The  oppositely charged black holes will not necessarily  annihilate, being pushed apart by the cosmic expansion,
or maybe they form  bound  systems stabilized  
by a scalar repulsion \cite{Herdeiro:2023mpt}. 
Such a mechanism for the charge generation should be verified, but if true,
it would imply that no new physics beyond the SM is  needed for monopoles. 
 Therefore, the extremal hairy EW black holes are 
the plausible candidates for magnetic  monopoles which
may perhaps exist in nature. 

Such black holes can show various properties, for example they 
should catalize the proton decay  \cite{Maldacena:2020skw}.  They can be detected when captured by 
a neutron star, causing a sudden change of the star's rotation period   \cite{Estes:2022buj}. 
Their phenomenology can be quite rich
and has been discussed in 
Refs.\cite{Bai:2020spd,Bai:2021ewf,Ghosh:2020tdu,Estes:2022buj,Diamond:2021scl}.

{\bf Acknowledgements.} We thank Eugen Radu, Maxim Chernodub, and Julien Garaud for discussions.






\vspace{1 cm}

\begin{center}
{\large\bf SUPPLEMENTAL MATERIAL}
\end{center}
We give below many technical details. 

\renewcommand{\thesection}{S.\arabic{section}}
\setcounter{equation}{0}

\renewcommand{\theequation}{\arabic{section}.\arabic{equation}}
\setcounter{equation}{0}

\section{DIMENSIONFUL AND DIMENSIONLESS QUANTITIES}
We denote dimensionful quantities by boldfaced letters and the dimensionless ones by ordinary lettters. 
The action of the bosonic part of the gravity-coupled electroweak theory 
is 
\be                                     \label{00}
{\bm {\mathcal S}}=\frac{1}{\bm c}\int\left(\frac{{\bm c}^4}{16\pi{\bm G}}\,{\bm R}+  {\bm  L}_{\rm WS} \right)\,\sqrt{-{\rm g}}\, d^4 {\bm x}\,,
\ee
where the electroweak Lagrangian is usually chosen in the form
\be                             \label{L}
{\bm L}_{\rm WS}&=&
-\frac{1}{4}\,{\bm \WW}^a_{\bm\mu\bm\nu}{\bm \WW}^{a\bm\mu\bm\nu}
-\frac{1}{4}\,{\bm\F}_{\bm\mu\bm\nu}{\bm\F}^{\bm\mu\bm\nu} \nn \\
&&-({\bm D}_{\bm\mu}{\bm \Phi})^\dagger {\bm D}^{\bm \mu}{\bm \Phi}
-{\bm \lambda}\left({\bm \Phi}^\dagger{\bm \Phi}-{\bm \Phi}_0^2\right)^2.~~~~~
\ee
Here ${\bm \Phi}_0$ is the Higgs field vacuum expectation value  and 
\be
{\bm \WW}^a_{\bm\mu\bm\nu}&=&\partial_{\bm\mu}{\bm \WW}^a_{\bm \nu}-\partial_{\bm\nu}{\bm \WW}^a_{\bm \mu}+
{\bm g}\epsilon_{abc}{\bm \WW}^b_{\bm \mu}{\bm \WW}^c_{\bm \nu},~~~~~~~\nn \\
{\bm\A}_{\bm\mu\bm\nu}&=&\partial_{\bm\mu}{\bm\A}_{\bm \nu}-\partial_{\bm\nu}{\bm\A}_{\bm \mu}\,,\nn \\
{\bm D}_{\bm\mu}{\bm \Phi}&=&\left(\partial_{\bm \mu}- \frac{i\bm g^\prime }{2}{\bm \A}_{\bm\mu}
- \frac{i\bm g}{2}\,\tau_a {\bm \WW}^a_{\bm\mu}
\right){\bm \Phi},
\ee
with $\partial_{\bm\mu}=\partial/\partial {\bm x^\mu}$. 
Denoting ${\bm g}_0=\sqrt{{\bm g}^2+{\bm g}^{\prime 2}}$ one can pass to dimensionless quantities by setting 
\be
{\bm g}&=&{\bm g}_0\,g,~~{\bm g}^\prime={\bm g}_0\,g^\prime,~~{\bm \WW}^a_{\bm \mu}=\frac{{\bm\Phi}_0}{g}\, \WW^a_\mu, \nn \\
~~{\bm \A}_{\bm \mu}&=&\frac{{\bm\Phi}_0}{g^\prime }\, \A_\mu,~~{\bm\Phi}={\bm\Phi}_0 \Phi,~~
{\bm\lambda}=\frac{\beta}{8}\,{\bm g}_0^2 \,,
\ee
while the spacetime coordinates 
${\bm x^\mu}={\bm l}_0 x^\mu $ with the length scale ${\bm l}_0=1/({{\bm g}_0{\bm\Phi}_0})$. The corresponding mass scale 
is ${\bm m}_0=\pmb{\hbar}/({\bm c\bm l}_0)=({\pmb\hbar}/{\bm c})\,{\bm g}_0{\bm\Phi}_0$. 

One has $g^\prime=\sin\thetaw\approx \sqrt{0.223}$ and $g=\cos\thetaw$. 
The dimensionful electron charge is ${\bm e}={\pmb{\hbar} \bm{c g}}_0\times e$ with 
$e\equiv gg^\prime=\sin\thetaw\cos\thetaw\approx0.416$ and the 
fine structure constant 
\be
\alpha&=&\frac{{\bm e}^2}{4\pi{\pmb{\hbar}\bm { c}}}=\frac{e^2}{4\pi}\,{\pmb{\hbar}\bm {cg}_0^2}\approx \frac{1}{137}~~~\Rightarrow~~~~\nn \\
\frac{1}{\bm {cg}_0^2}&=&\frac{e^2}{4\pi\alpha}\,\pmb{\hbar}=1.89\,{\pmb\hbar},~~~
\frac{1}{\bm{g}_0}=\frac{\pmb{\hbar}\bm{c}}{\bm e}\,e=\frac{e}{4\pi\alpha}\,\bm{e}.
\ee
Injecting everything into \eqref{00} and introducing the dimensionless gravitational coupling,
\be            \label{kappa}
\kappa=\frac{8\pi {\bf G {\bm \Phi}_0^2}}{{\bm c}^4},
\ee
the action becomes 
$
{\bm {\mathcal S}}={\pmb\hbar}\times {\mathcal S}
$
with 
\be                                     \label{II}
{\mathcal S}=\frac{e^2}{4\pi\alpha}\int \left(\frac{1}{2\kappa}\,R+{\cal L}_{\rm WS}
\right)\sqrt{-\rm g}\, d^4x,~~~~
\ee
and
\be                \label{LEW}
{\cal L}_{\rm WS}=-\frac{1}{4g^2}\,\WW^a_{\mu\nu}\WW^{a\mu\nu}
-\frac{1}{4g^{\prime 2}}\,{\F}_{\mu\nu}{\F}^{\mu\nu} \nn \\
-(D_\mu\Phi)^\dagger D^\mu\Phi
-\frac{\beta}{8}\left(\Phi^\dagger\Phi-1\right)^2.
\ee
This is the action in 
Eq.\eqref{I}
in the main text.

The Z-boson mass is $\mz=1/\sqrt{2}$ whose  dimensionful version 
\be
{\bm \mz}=\frac{{\bm m}_0}{\sqrt{2}}=\frac{\pmb{\hbar}\,{\bm{g}_0{\bm \Phi}_0 } }{\sqrt{2}{\bm c}}=91.18 \,{\rm GeV}/{{\bm c}^2}
\ee
determines 
the mass scale  ${\bm m}_0=128.9~{\rm GeV}/{\bm c}^2$, and the length scale ${\bm l}_0=1.53\times 10^{-16}~{\rm cm}$,
hence 
\be            \label{kappa1}
\kappa
=\frac{4\,e^2}{\alpha}
\left(\frac{{\bm m}_{\mbox{\tiny Z}}}{{\bf M}_{\rm Pl}}\right)^2=5.30\times 10^{-33}. 
\ee

The dimensionful version of the  ADM mass is determined from the asymptotic form of the ${\rm g}_{00}$ metric 
coefficient, 
\be               \label{Ninf}
 -g_{00}=1-\frac{2M}{\rr}+\ldots = 1-\frac{2{\bf GM}}{\bm{c}^2\bf{r}}+\ldots\,,
 \ee
 where ${\bm r}={\bm l}_0\, r$ hence 
 \be           \label{M1}
\frac{\bf M}{\bm{m}_0}&=&\frac{\bm{c}^2\bm{l}_0 }{{\bf{G}}\bm{m}_0}\, M=
 \frac{8\pi}{\kappa}\times \frac{M}{\pmb{\hbar}\,\bm{c g}_0^2} \nn \\
 &=&
  \frac{e^2}{4\pi\alpha}\times
 \frac{8\pi}{\kappa}\,M\equiv \frac{e^2}{4\pi\alpha}\times
 {\cal M}. 
 \ee
 
 \section{FIELD EQUATIONS} 
 \setcounter{equation}{0}
Varying the action \eqref{II} with respect to the EW fields 
gives the equations,
\begin{align}           
\nabla^\mu {B}_{\mu\nu}&=g^{\prime 2}\,\frac{i}{2}\,
(\Phi^\dagger D_\nu\Phi -(D_\nu\Phi)^\dagger\Phi
),\nn 
\\
{\cal D}^\mu \WW^a_{\mu\nu}
&=g^{2}\,\frac{i}{2}\,
(
\Phi^\dagger\tau^a D_\nu\Phi
-(D_\nu\Phi)^\dagger\tau^a\Phi
)
, \nn 
\\
D_\mu D^\mu\Phi&-\frac{\beta}{4}\,(\Phi^\dagger\Phi-1)\Phi=0,      \label{P2}
\end{align}
with ${\cal D}_\mu\WW^a_{\alpha\beta}=\nabla_\mu \WW^a_{\alpha\beta}
+\epsilon_{abc}\WW^b_\mu\WW^c_{\alpha\beta}$ where $\nabla_\mu$ is the geometrical covariant derivative with respect 
to the spacetime metric ${\rm g}_{\mu\nu}$. 
Varying the action with respect to the latter yields the Einstein equations 
\be             \label{Einst}
G_{\mu\nu}=\kappa\, T_{\mu\nu} \,,
\ee
with  the energy-momentum tensor
\be                      \label{TT}
T_{\mu\nu}&=&
\frac{1}{g^2}\,\WW^a_{~\mu\sigma}\WW^{a~\sigma}_{~\nu}
+\frac{1}{g^{\prime\,2}}B_{\mu\sigma}B_\nu^{~\sigma}  \\
&&+(D_\mu\Phi)^\dagger D_\nu\Phi \nn 
+(D_\nu\Phi)^\dagger D_\mu\Phi
+{\rm g}_{\mu\nu}\mathcal{L}_{\rm WS}\,. \nn 
\ee
The vacuum is defined as the configuration with $T_{\mu\nu}=0$. Modulo gauge transformations, it can be chosen as 
\be
\WW^a_\mu=\A_\mu=0,~~~~~~ 
 \Phi=\begin{pmatrix}
0  \\
1
\end{pmatrix},~~~~~~~{\rm g}_{\mu\nu}=\eta_{\mu\nu},~~~~~
\ee 
where $\eta_{\mu\nu}={\rm diag}[-1,1,1,1]$ is the Minkowski metric. 
Allowing for small fluctuations around the vacuum and 
linearizing the field equations 
with respect to the fluctuations, gives the perturbative mass spectrum 
containing the massless photon, massless graviton, 
and the massive Z, W and Higgs bosons with dimensionless masses
\be                                   \label{masses}
\mz=\frac{1}{\sqrt{2}},~~~
\mw=g\,\mz,~~~
\mh=\sqrt{\beta}\,\mz.
\ee

 \section{GAUGE TRANSFORMATIONS}
 \setcounter{equation}{0}

 The action \eqref{II} is invariant under 
SU(2)$\times$U(1) gauge transformations
\be                               \label{gauge}
\Phi\to {\rm \U}\,\Phi,~~~
{\cal W}\to {\rm \U}\,{\cal W}\,{\rm \U}^{-1}
+i\,{\rm \U}\,\partial_\mu {\rm \U}^{-1}dx^\mu\,,
\ee
with 
\be                            \label{U}
{\cal W}=
\frac12\, (B_\mu+\tau^a\WW^a_\mu)\, dx^\mu\,,~~
{\rm \U}=\exp\left(\frac{i}{2}\,\Sigma+\frac{i}{2}\,\tau^a\theta^a\right), \nn
\ee
where $\Sigma$ and $\theta^a$ are functions of $x^\mu$. 

The ansatz for the electroweak fields in Eq.\eqref{metr} in the main text is
\be               \label{RR}
\WW&=&\T_2\left( F_1\,d\rr+F_2\,d\vartheta \right)+\nu \left( \T_3\,F_3-\T_1 F_4\,\right)d\varphi\,, \nn \\
B&=&\nu\, Yd\varphi\,,~~~~~~
\Phi=\begin{pmatrix}
\phi_1 \\
\phi_2
\end{pmatrix}\,,
\ee
with $\nu=-n/2$. It 
keeps its form under gauge 
transformations generated by ${\rm \U}=\exp\left\{i\chi(\rr,\vartheta)\T_2\right\}$, whose effect  is 
\be               \label{res}
&&F_1\to F_1+\partial_\rr \chi,~~
F_2\to F_2+\partial_\vartheta \chi,~~Y\to Y,\nn \\
&&F_3\to F_3\,\cos\chi-F_4\,\sin\chi,~~~~~~\nn \\
&&F_4\to F_4\,\cos\chi+F_3\,\sin\chi\,,~~~~~~\nn \\
&&\phi_1\to \phi_1\,\cos(\chi/2)+ \phi_2\,\sin(\chi/2),~~\nn \\
&&\phi_2\to \phi_2\,\cos(\chi/2)- \phi_1\,\sin(\chi/2).
\ee
The fields in \eqref{RR} are singular at the symmetry axis. To remove the singularity, one sets 
\be                \label{var}
&&F_1=-\frac{H_1(\rr,\vartheta)}{\rr\sqrt{\rm N}},~~~~~~~F_2=H_2(\rr,\vartheta),~~\nn \\
&&F_3=\cos\vartheta+H_3(\rr,\vartheta)\sin\vartheta\,,~~~
F_4=H_4(\rr,\vartheta)\sin\vartheta\,,~~\nn \\
&&Y=\cos\vartheta+y(\rr,\vartheta)\sin\vartheta\,. 
\ee
The gauge transformation 
generated by
\be                   \label{Ureg} 
 {\rm \U}_\pm =
 e^{- i\nu\varphi \T_3 }
 e^{- i\vartheta \T_2}
 e^{\pm i\nu\varphi/2}
  \ee
brings the SU(2) field to the form 
 \be           \label{gauge2}
\WW&=&\T_\varphi\left(-\frac{H_1}{\rr\sqrt{\N}}\,d\rr+(H_2-1)\,d\vartheta \right) \nn \\
&&+\nu \,\left( \T_r\,H_3+\T_\vartheta\,(1-H_4) \right)\sin\vartheta\, d\varphi\,,
\ee
where 
$
\T_r=\T_1\sin\vartheta\cos(\nu\varphi)+\T_2\sin\vartheta\sin(\nu\varphi)+\T_3\cos\vartheta
$
and $\T_\vartheta=\partial_\vartheta \T_r$, $\T_\varphi=\partial_\varphi \T_r/({\nu\sin\vartheta})\,$. 
This field is $\varphi$-dependent but  regular at the symmetry axis, owing to the boundary conditions 
defined in Eq.\eqref{bc} below
(see \cite{GVII} for details). 
The U(1) and Higgs fields become 
\be          \label{gauge2a}
B_\pm &=&\nu\,(\cos\vartheta\pm 1+\,y\,\sin\vartheta)\, d\varphi\,, \\
\Phi_\pm&=&
e^{\pm i\nu\varphi/2}\,
\begin{pmatrix}
\left.\left. \,\right(\phi_1\,\cos(\vartheta/2)-\phi_2\,\sin(\vartheta/2)\right) e^{-i\nu\varphi/2} \\
\left.\left. \,\right(\phi_1\,\sin(\vartheta/2)+\phi_2\,\cos(\vartheta/2)\right) e^{+i\nu\varphi/2}
\end{pmatrix}, \nn 
\ee
where the upper $``+"$ and lower $``-"$ signs correspond to two local gauges used, respectively, 
in the southern and northern hemi-spheres. These two gauges  are related to each other in the equatorial region 
by the U(1) transformation with ${\rm \U}=\exp(i\nu\varphi)$ (see \cite{GVII} for details). 
When expressed in these two local gauges, the U(1) and Higgs fields are everywhere regular. 

The above formulas suggest that $\nu=-n/2$ should be integer. However, it can be half-integer as well. In that case 
the $\varphi$-dependent part of $\T_r$ changes sign under $\varphi\to\varphi+2\pi$, but the correct sign can be restored 
by the global gauge transformation generated by ${\rm \U}=\tau_3$ or ${\rm \U}=-\tau_3$ whose effect is 
\be            \label{TTT}
\T_1\to -\T_1,~~~~~\T_2\to -\T_2\,, 
\ee
so that the field \eqref{gauge2} does not change. Similarly, if $\nu$ is half-integer then the lower component of $\Phi_{+}$ and the 
upper component of $\Phi_{-}$ change sign under  $\varphi\to\varphi+2\pi$, but the correct sign can be restored
by the global gauge transformation generated by ${\rm \U}=\tau_3$ and ${\rm \U}=-\tau_3$, respectively. 
The fields \eqref{gauge2a} then remain invariant. As a result, all integer values of $n$ are allowed.

\section{DIFFEOMORPHISMS}
 \setcounter{equation}{0}

The line element  in Eq.\eqref{metr} of the main text 
\be                   \label{metr11}
ds^2&=&-e^{2\Uo} {\rm N}(\rr)\, dt^2+e^{-2\Uo} dl^2, \nn \\
dl^2&=&
e^{2\Ko}\left[\frac{d\rr^2}{{\rm N}(\rr)}+\rr^2d\vartheta^2\right]+e^{2 \So}\,\rr^2\sin^2\vartheta d\varphi^2,
\ee
keeps its form under diffeomorphisms $\rr\to\tilde{\rr}=\tilde{\rr}(\rr)$ whose effect  is 
$\Uo\to \tilde{\Uo}$, $\N\to\tilde{\N}$, $\Ko\to\tilde{\Ko}$, $\So\to\tilde{\So}$, where 
\be
\frac{d\rr}{\rr \sqrt{\N}}=\frac{d\tilde{\rr}}{\tilde{\rr} \sqrt{\tilde{\N}}},~~~~~~~~~e^{\Uo}\sqrt{\N}=e^{\tilde{\Uo}}\sqrt{\tilde{\N}},\nn \\
\rr\sqrt{\N} e^{\Ko}=\tilde{\rr}\sqrt{\tilde{\N}} e^{\tilde{\Ko}},~~~~~~~~~
\rr\sqrt{\N} e^{\So}=\tilde{\rr}\sqrt{\tilde{\N}} e^{\tilde{\So}}.
\ee
This symmetry can be fixed by making a specific choice for the auxiliary function $\N(\rr)$, for example one can set $\N(\rr)=1$. 
We choose  $\N(\rr)=1-\rrh/\rr$ for the non-extremal solutions, then $\rr=\rrh$ 
corresponds to the position of the event horizon. However, since $\rr$ is not 
the Schwarzschild coordinate, the parameter $\rrh$ does not have a direct physical meaning, although its value determine the 
 horizon area. One has $\rrh\in [0,\rrh^0]$ and when $\rrh\to 0$, the solutions approach the extremal limit with a non-zero horizon area, 
while for $\rrh\to \rrh^0$ they merge with the RN solution 
and one has $\rrh^0=r_{+}-r_{-}$ where $r_\pm=M\pm \sqrt{M^2-Q^2}$ are the two RN horizons. 

For the extremal solutions we make the choice  $\N(\rr)={\rm k}(\rr)\times(1-\rrh/\rr)^2$
with 
\be              \label{k_ex}
{\rm k}(\rr)=1-\left.\left.\frac{\Lambda}{3}\right[\rr^2+2\,\rr_{\rm H}\,\rr+3\,\rr_{\rm H}^2\right]\times 
\frac{1+\rr_{\rm H}^4}{1+\rr^4}\,.
\ee
The horizon is again at $\rr=\rrh$  and it is degenerate so that the surface gravity vanishes. 
The parameter $\rrh$ again does not have a direct physical meaning and is not related anymore to the horizon area, 
so that, for example,  one can set $\rrh=1$. 
However,  for the extremal solutions in phase I which have a spherical horizon  
it is convenient to choose $\rrh=r_{\rm ex}$ where $r_{\rm ex}$ is the radius of the extremal RNdS solution 
described by Eq.\eqref{RNdS} in the main text. In this case the radial coordinate $\rr$ coincides at the horizon  with the Schwarzschild coordinate $r$ and the 
function $\N(\rr)$ has the same limit as $N(r)$ for the extremal RNdS solution. For extremal solutions in phase II one can  set $\rrh=const$.

\section{CHARGE AND CURRENTS}
\setcounter{equation}{0}

The 
electromagnetic and Z fields are defined according to Nambu \cite{Nambu:1977ag},
\be                                  \label{Nambu}
\FF_{\mu\nu}&=&\frac{g}{g^\prime}\,  
\A_{\mu\nu}-\frac{g^{\prime}}{g}\,n^a\WW^a_{\mu\nu}\,,~~~~~~\nn \\
{\mathcal Z}_{\mu\nu}&=&\A_{\mu\nu}+n^a\WW^a_{\mu\nu}\,,
\ee
where 
$
n^a=\Phi^\dagger\tau^a\Phi/(\Phi^\dagger\Phi). 
$
 The 2-form 
  \be
 \mathcal{F}=\frac12 {\cal F}_{\mu\nu} dx^\mu\wedge dx^\nu\,,
 \ee
 is closed in the Higgs vacuum where ${\cal F}_{\mu\nu}$ reduces to the usual electromagnetic field tensor. 
 Away from the Higgs vacuum the form ${\cal F}$ is not necessarily closed since in this case there is no reason for Maxwell equations to apply. 
 Defining the dual field tensor, 
 \be
\tilde{\FF}^{\mu\nu}=\frac{1}{2\sqrt{-\rm g}}\,\epsilon^{\mu\nu\alpha\beta}\FF_{\alpha\beta}\,,~~~~~
\ee
the conserved electric current and magnetic densities can be defined as 
 \be             \label{cur}
 \J^\mu &=&\frac{1}{4\pi}\,\frac{1}{\sqrt{-\rm g}}\,\partial_\nu \left(\sqrt{-\rm g}\,\FF^{\mu\nu}\right). \\
 \tilde{\J}^\mu &=&\frac{1}{4\pi}\,\frac{1}{\sqrt{-\rm g}}\,\partial_\nu \left(\sqrt{-\rm g}\,\tilde{\FF}^{\mu\nu}\right). 
 \ee
 The flux of $\mathcal{F}$ through a two-sphere $S^2$ at spatial infinity defines the conserved magnetic charge, 
 \be
 P=\frac{1}{4\pi}\oint_{S^2}  {\cal F}. 
  \ee
Since $\FF_{\mu\nu}$ consists of two parts 
  determined by the contribution of $\A_{\mu\nu}$ and of $\WW^a_{\mu\nu}$, 
  the magnetic charge splits, 
 \be                   \label{PT}
 P= P_{\rm U(1)}+P_{\rm SU(2)}, ~~~
 \ee
 with 
 \be               \label{PB}
 P_{\rm U(1)}&=&\frac{1}{4\pi}\,\frac{g}{g^\prime}\oint_{S^2} \frac12 B_{\mu\nu} dx^\mu \wedge dx^\nu \\
 &=&\frac{1}{4\pi}\,\frac{g}{g^\prime}\, \oint_{S^2} dB\,=\frac{g}{g^\prime}\frac{n}{2}=g^2\, \frac{n}{2e}=g^2 P, \nn
\ee
 where we used the expression for the $B$-field in \eqref{RR},\eqref{var} and the boundary conditions 
 specified below in \eqref{bc}. 
 The integral here does not depend on the radius of the sphere hence 
 this part of the charge is always contained 
 inside the black hole. 
 
 The remaining part of the charge, 
 \be
 P_{\rm SU(2)}=P-P_{\rm U(1)}=g^{\prime 2} P, 
 \ee
 is not always contained 
 inside the black hole. Its  part distributed outside can be obtained by subtracting the flux at the horizon from flux at infinity,
  \be                 \label{Phh}
4\pi  P_{\rm h}&=&\oint_{\rr\to\infty} \mathcal{F} -\oint_{\rr=\rrh} \mathcal{F} 
=\int_{\rr>\rrh} d\mathcal{F}   \\
&=& \int_{\rr>\rrh} \frac12\, \epsilon^{ijk}\partial_i \mathcal{F}_{jk}\, d^3x= \int_{\rr>\rrh} \tilde{\mathcal{J}}^0\sqrt{\rm -g}\, d^3x. \nn
 \ee
 Since the form $\mathcal{F}_{\mu\nu}$ is not necessarily closed, 
 the latter integral can be non-zero. 
 One has 
 \be               \label{FF} 
 \mathcal{F}_{\mu\nu}=F_{\mu\nu}+e\psi_{\mu\nu}\,,
 \ee
 where $F_{\mu\nu}=\partial_\mu A_\nu-\partial_\nu A_\mu$ is a closed form while the rest is not closed, 
 \be            \label{Hooft}
 g^2\,\psi_{\mu\nu}=-\epsilon_{abc} n^a\mathcal{D}_\mu n^b\mathcal{D}_\nu n^c\,,
 \ee
 with  $\mathcal{D}_\mu n^a=\partial_\mu n^a+\epsilon_{abc} \WW^b_\mu n^c$. 
 This tensor is made of massive fields and approaches zero exponentially fast  at infinity,
 therefore, 
  \be
4\pi  P_{\rm h} &=&e\int_{\rr>\rrh} \frac12\,\epsilon^{ijk}\,\partial_i \mathcal{\psi}_{jk} \,d^3x\nn \\
&=&-e\oint_{S^2,\rr=\rrh} \frac12\, \psi_{ik}\, dx^i\wedge dx^k. 
\ee

The tensor $F_{\mu\nu}$ leads to another definition of the electric current, 
\be             \label{cur1}
 J^\mu &=&\frac{1}{4\pi}\,\frac{1}{\sqrt{-\rm g}}\,\partial_\nu \left(\sqrt{-\rm g}\,F^{\mu\nu}\right).
  \ee
  This definition is better adapted for the perturbative analysis carried out in Section S.9 below.
   Integrating the current densities $\mathcal{J}^\mu$ and $J^\mu$ over a 2-surface gives currents. 
 
 In the static case, denoting $\mathcal{N}=\sqrt{-\rm g_{00}}$, the relation \eqref{cur} (similarly for \eqref{cur1}) can 
 be represented in the form language as 
 \be
 d(\mathcal{N}\ast\FF)=4\pi \mathcal{N}\ast \mathcal{J}
 \ee
 where $\ast$ denotes the Hodge duality operator on the spacelike 3-surface. 
 Both sides of this relation are 2-forms. Integrating over a 2-surface $\Sigma$ and applying Green's theorem yields 
 \be
\frac{1}{4\pi} \int_{\partial \Sigma} \mathcal{N}\ast\FF =\int_\Sigma \mathcal{N}\ast \mathcal{J}\equiv \mathcal{I}.
 \ee
Here on the left is the circulation of the magnetic field along the boundary $\partial\Sigma$, hence on the right is the 
total current through $\Sigma$. 

In the situation considered in the main text the currents are along the azimuthal direction, 
hence they flow through planes of constant $\varphi$. 
The above definition then implies that the total azimuthal current in the upper half-space, $\vartheta<\pi/2$, is 
\be
\mathcal{I}_{+}=\int_{\rrh}^\infty d\rr \int_0^{\pi/2} d\vartheta \, \sqrt{-\rm g}\, \J^\varphi;
\ee
similarly for the current $I_{+}$ associated with $J^\varphi$. 
 A similar integral with $\vartheta\in[\pi/2,\pi]$ yields the current $\mathcal{I}_{-}$ in the lower half-space. 
 Since $\mathcal{J}^\varphi$ is antisymmetric with respect to the equatorial plane, the total current is zero, 
  $\mathcal{I}_{+}+\mathcal{I}_{-}=0$. 
 The dimensionful current expressed in amperes is
 \be
 {\bm I}={\bm c\bm \Phi}_0\, \mathcal{I}=\frac{e}{4\pi\alpha}\, \frac{\bm e}{\bm{t}_0}\,\mathcal{ I}=1.42\times10^8\times\mathcal{ I} \,\,\text{A},
 \ee
 where ${\bm t}_0={\bm l}_0/{\bm c}=5.1\times 10^{-27}$ sec.

  \section{BOUNDARY CONDITIONS AND NUMERICAL PROCEDURE}
 \setcounter{equation}{0}
 The axially symmetric EWS fields are the line element in \eqref{metr11} 
 and the electroweak fields in \eqref{RR},\eqref{var}
 expressed in terms of 10 functions $\Uo,\Ko,\So,H_1,H_2,H_3,H_4,y,\phi_1,\phi_2$
 depending on $\rr,\vartheta$. 
 The fields in \eqref{RR},\eqref{var} have a residual U(1) gauge invariance  \eqref{res} that we fix by imposing 
the gauge condition $\rr\sqrt{\N}\partial_\rr H_1=\partial_\vartheta H_2$. This gives 10 elliptic equations 
whose solutions  fulfil also  the first order gravitational constraints  owing to the 
boundary conditions described by Eq.\eqref{bc} below.

 We use 
 the compact radial variable $\rm x\in[0,1]$ such that 
$
\rr=\sqrt{\rrh^2+{\rm x}^2/(1-{\rm x})^2}
$
and assume the reflection invariance of the energy density under $\vartheta\to \pi-\vartheta$ to restrict 
the range to $\vartheta\in[0,\pi/2]$. 
The following 
boundary conditions at the borders of the 
integration domain are imposed:
\be          \label{bc}
\underline{\text{axis}, \vartheta=0}: &&
~\partial_\vartheta=0~ \text{for}~  \Uo,\So,H_2,H_4,\phi_2;\nn  \\
&&~\Ko-\So=H_1=H_3=y=\phi_1=0;  \nn \\
\underline{\text{equator}, \vartheta=\pi/2}: &&
~\partial_\vartheta=0 \text{~for~}  \Uo,\Ko,\So,H_2,H_4,\phi_2; \nn \\ 
&&~H_1=H_3=y=\phi_1=0;  \nn \\
\underline{\text{horizon}, {\rm x}=0}:&& ~H_1=0:  \partial_{\rm x}=0 \text{~for the rest} \nn \\
\underline{\text{infinity}, {\rm x}=1}: &&~\phi_2=1; \text{~0 for the rest}. 
\ee

It turns out that these boundary conditions automatically imply that at the symmetry axis one has $H_2=H_4$ and 
$\partial_\vartheta\Ko=0$, as needed for the regularity. 

We solve the field equations with these boundary conditions to determine the components of the ``state vector''
\be       \label{state}
\Psi=[H_1,H_2,H_3,H_4,y,\phi_1,\phi_2,\Uo,\Ko,\So],
\ee
which are functions of $\rr,\vartheta$. We solve the equations with the 
FreeFem  numerical solver based on the finite element method  \cite{MR3043640}. 
This solver uses the weak form of  differential equations obtained by transforming them into integral equations, 
expanding with respect to basis functions obtained by triangulating  the integration domain, and handling 
the non-linearities with the Newton-Raphson procedure. 
The quality of the solutions is checked by monitoring the 
virial relation that should hold on-shell, 
\be            \label{vir}
v \equiv \int_{\rr>\rr_{\rm H}} T^\mu_{~\nu}\nabla_\mu\zeta^\nu \,\sqrt{\rm -g} \,d^3x=0,
\ee
where $\zeta_\mu dx^\mu= \rr\,e^{\Ko-\Uo}\, d\rr$ and $T^\mu_{~\nu}$ is the EW energy-momentum tensor \eqref{TT}. 
For a  domain triangulation by $100\times 60$ elements we typically obtain  $v\sim 10^{-7}$.

 \section{HORIZON MASS}
 \setcounter{equation}{0}

 The horizon radius of the RN solution with the magnetic charge $P$ and mass $M$ is $r_{h}=M+\sqrt{M^2-Q^2}$ 
 where $Q^2=(\kappa/2) P^2$, hence 
 \be            \label{MM}
 M=\frac{r_{h}}{2}+\frac{Q^2}{2 r_{h}}=\frac{r_{h}}{2}+\frac{\kappa P^2}{4 r_{h}}.
 \ee
 On the other hand, 
the total magnetic charge consists of the U(1) part and SU(2) part, 
$
P=P_{\rm U(1)}+P_{\rm SU(2)}\,
$
with $P_{\rm U(1)}=g^2 P$ and $P_{\rm SU(2)}=g^{\prime 2} P$ 
and one has 
\be               \label{P22}
P^2=\frac{1}{g^2}\, P_{\rm U(1)}^2+\frac{1}{g^{\prime 2}}\, P_{\rm SU(2)}^2.
\ee
This formula is important: although $P$ is the sum of two charges, $P^2$ is not the square of the sum
because the two charges are of different nature 
and only the square of each charge contributes to the energy.

The magnetic charge of a hairy black hole splits as 
$
P= P_{\rm H}+P_{\rm h}\,
$
where $P_{\rm H}$ is the charge contained inside the black hole and 
$P_{\rm h}$ is the charge contained outside in the hair defined by \eqref{Phh}. 
The hair charge is always a part of the non-Abelian charge $P_{\rm SU(2)}$, hence 
\be
P_{\rm h}=\tilde{\lambda} P_{\rm SU(2)},~~~~P_{\rm H}= P_{\rm U(1)}+(1-\tilde{\lambda})P_{\rm SU(2)},
\ee
where the parameter $\tilde{\lambda}$ varies from zero for black holes which bifurcate with RN till unity for 
the extremal  black holes in phase I. Comparing with \eqref{P22}, the square of the charge contained inside the 
black hole  is computed according the rule   
\be               \label{P222}
P_{\rm H}^2&=&\frac{1}{g^2}\, P_{\rm U(1)}^2+\frac{(1-\tilde{\lambda})^2}{g^{\prime 2}}\, P_{\rm SU(2)}^2 \nn \\
&=&
(g^2+(1-\tilde{\lambda})^2g^{\prime 2})P^2.
\ee
Therefore, {\it defining} the size of a hairy black hole with a horizon area ${\rm A}_{\rm H}$ as  $r_h=\sqrt{{\rm A}_{\rm H}/(4\pi)}$
and comparing with \eqref{MM}, the horizon mass is 
\be            \label{MMM}
 M_{\rm H}=\frac{r_{h}}{2}+\frac{\kappa P^2_{\rm H}}{4 r_{h}}=
 \frac{r_{h}}{2}+\frac{\kappa P^2}{4 r_{h}}(g^2+(1-\tilde{\lambda})^2g^{\prime 2}).~~~~
 \ee
 For the hairy solutions bifurcating with the RN branch one has $\tilde{\lambda}=0$ and the formula reduces to \eqref{MM}. 
 For the extremal hairy solutions in phase I one has $\tilde{\lambda}=1$ and 
 \be            
 M_{\rm H}=
 \frac{r_{h}}{2}+\frac{\kappa g^2\,P^2}{4 r_{h}}=\frac{r_{h}}{2}+g^2\,\frac{Q^2}{2 r_{h}}.
 \ee
 Notice finally that $P_{\rm h}=\tilde{\lambda} P_{\rm SU(2)}=\tilde{\lambda} g^{\prime 2} P\equiv \lambda P$
 hence $\tilde{\lambda}=\lambda/g^{\prime 2}$ where $\lambda=P_{\rm h}/P$ is used in the main text. 
 
\section{QUADRUPOLE MOMENTS}
\setcounter{equation}{0}

Far away from the horizon all massive fields approach vacuum values and one has 
\be
B_\mu=W^3_\mu\equiv eA_\mu,~~~W^1_\mu=W^2_\mu=0, ~~~ \Phi=\begin{pmatrix}
0  \\
1
\end{pmatrix},~~~
\ee
hence the Einstein-Weinberg-Salam theory reduces  to the electrovacuum, 
\be
{\cal L}=\frac{1}{2\kappa}\, R-\frac{1}{4 }\, F_{\mu\nu}F^{\mu\nu},
\ee
where $F_{\mu\nu}=\partial_\mu A_\nu-\partial_\nu A_\mu$. 
The  solutions approach at large distances 
the magnetic RN  configuration of mass $M$ and charge $Q=\sqrt{\kappa/2}\,P$. 
Choosing the gauge where $\N(\rr)=1$, the line element in \eqref{metr11}  becomes 
\be              \label{ds4}
ds^2&=& -e^{2\Uo} dt^2+e^{-2\Uo} dl^2,~~ \\
dl^2&\equiv &h_{ik}dx^i dx^k=e^{2\Ko}(d\rr^2+\rr^2 d\vartheta^2)+\rr^2 e^{2\So}\sin^2\vartheta\,d\varphi^2\,, \nn
\ee
while the purely magnetic Maxwell field $F_{\mu\nu}$ can be expressed in terms of a magnetic potential $\Psi$ as 
\be            \label{dual} 
\sqrt{\frac{\kappa}{2}}\,F_{ik}=e^{-2\Uo} \sqrt{h} \,\epsilon_{iks}\, \partial^s\Psi~\,.
\ee
Our solutions in the far field region reduce to
\be                   \label{metr1}
e^{2\Uo}&=&e^{2\delta \Uo}\frac{\X^2-\mu^2}{(\X+M)^2},~~~~
e^{\Ko}=e^{\delta \Ko}\left(1-\frac{\mu^2}{4\xx^2}\right),\nn \\
e^{\So}&=&e^{\delta \So}\left(1-\frac{\mu^2}{4\xx^2}\right),~~~~~
\Psi=\frac{Q}{\X+M}+\delta\Psi,~
\ee
with $x=\rr+\mu^2/(4\rr)$ and $\mu^2=M^2-Q^2$. In the leading order this is the RN solution 
in the isotropic coordinates, while 
the subleading terms describe deviations from the  RN background, 
\be     \label{damRN}
\delta \So&=&-\frac{s_2}{2\xx^2}+\ldots,   \nn \\
\delta \Ko&=&s_2\,\frac{2\sin^2\vartheta-1}{2\xx^2}+\ldots , \nn \\
\delta \Uo&=&\frac{Q_g\,(1-3\cos^2\vartheta)-Ms_2/3 }{2\xx^3}+\ldots , \nn \\
\delta \Psi&=&\frac{Q_m\,(1-3\cos^2\vartheta)+Qs_2/3 }{2\xx^3}+\ldots  ,
\ee
where $Q_g,Q_m,s_2$ are integration constants and the dots denote higher order  terms. 
The values of these constants  can be determined  from our numerical solutions, 
taking into account that the latter  are obtained in a different radial gauge where $\N(\rr)\neq 1$.

Introducing the Ernst potential, 
\be               \label{Ernst}
{\cal E}=e^{2\Uo}-\Psi^2,
\ee
the gravitational $Q_{\rm G}$ and magnetic $Q_{\rm M}$ quadrupole moments defined within the Geroch-Hansen 
approach 
\cite{Geroch:1970cd,Hansen} are determined by the behaviour of 
\be          
\xi=\frac{1-{\cal E}}{1+{\cal E}},~~~~~q=\frac{2\Psi}{1+{\cal E}}
\ee
near spatial infinity. This requires  \cite{Fodor,Fodor1} 
passing to the Weyl coordinates where 
\be
dl^2&=&e^{2K(\rho,z)}(d\rho^2+dz^2)+\rho^2 d\varphi^2,
\ee
(the function $K$ can be expressed in terms of $\Ko,\So$) 
and considering the asymptotic expansions at the symmetry axis, $\rho=0$,  for $z\to\infty$,
\be            \label{GH}
\xi=\sum_{k\geq 0} \frac{a_k}{z^{k+1}},~~~~~~
q=\sum_{k\geq 0} \frac{b_k}{z^{k+1}},  
\ee
from where 
$Q_{\rm G}=-a_2$, $Q_{\rm M}=-b_2$. 
Transforming the 3-metric in \eqref{ds4} to the Weyl coordinates, computing the potentials $\xi$ and $q$ 
and expanding them according to \eqref{GH} yields 
\be
Q_{\rm G}=-Q_g-\frac23\, M\, s_2,~~~Q_{\rm M}=Q_m-\frac23\, Q\, s_2.~~~
\ee

\section{BEYOND AXIAL SYMMETRY}
\setcounter{equation}{0}

Although this is not the main subject of this work, we consider also 
non-axially symmetric solutions by adapting  the perturbative  approach of \cite{Ridgway:1995ke}. 
The idea is to expand around the RN solution up to the second order terms and then minimize the energy. 

Passing to the unitary gauge where the Higgs field $\Phi^{\rm tr}=(0,\phi)$ with real-valued $\phi$, 
defining the complex-valued vector field $w_\mu=(\WW^1_\mu+i\WW^2_\mu)/g$ with the strength 
$
w_{\mu\nu}=D_\mu w_\nu-D_\nu w_\mu
$
where $D_\mu=\nabla_\mu +i\WW^3_\mu$, 
the EW Lagrangian \eqref{LEW} assumes the form 
\be           \label{LEW1}
-{\cal L}_{\rm EW}=\frac{1}{4g^{\prime 2}}(B_{\mu\nu})^2+\frac{1}{4g^{2}}(\WW^3_{\mu\nu})^2
+ \frac{1}{4}\,|w_{\mu\nu}|^2 ~~~~~~~~~ \\
+(\partial_\mu\phi)^2+\frac14\left( g^2|w_\mu|^2+(B_\mu-\WW^3_\mu)^2 \right)\phi^2+\frac{\beta}{8}(\phi^2-1)^2.~~\nn
\ee
Here $B_{\mu\nu}=\partial_\mu B_\nu-\partial_\nu B_\mu$ 
and 
\be                \label{psi}
\WW^3_{\mu\nu}&=&\partial_\mu \WW^3_\nu-\partial_\nu \WW^3_\mu+ g^2\psi_{\mu\nu},\nn \\
\psi_{\mu\nu}&=&\frac{i}{2}(w_\mu\bar{w}_\nu-w_\nu\bar{w}_\mu). 
\ee
The electromagnetic and Z-fields \eqref{Nambu} then become 
\be
e\FF_{\mu\nu}=g^2 B_{\mu\nu}+g^{\prime 2}\WW^3_{\mu\nu}&\equiv& e F_{\mu\nu}+ e^2\psi_{\mu\nu}\,,\nn \\
{\mathcal Z}_{\mu\nu}=B_{\mu\nu}-\WW^3_{\mu\nu}&\equiv&  Z_{\mu\nu}- g^2\psi_{\mu\nu}\,,
\ee
where 
$F_{\mu\nu}=\partial_\mu A_\nu-\partial_\nu A_\mu$ and 
$Z_{\mu\nu}=\partial_\mu Z_\nu-\partial_\nu Z_\mu$ 
with 
\be
eA_\mu=g^2\,B_\mu+g^{\prime 2}\, \WW^3_\mu,~~~~
Z_\mu=B_\mu-\WW^3_\mu\,.
\ee
\begin{figure}[!b]
    \centering
        \includegraphics[scale=0.5]{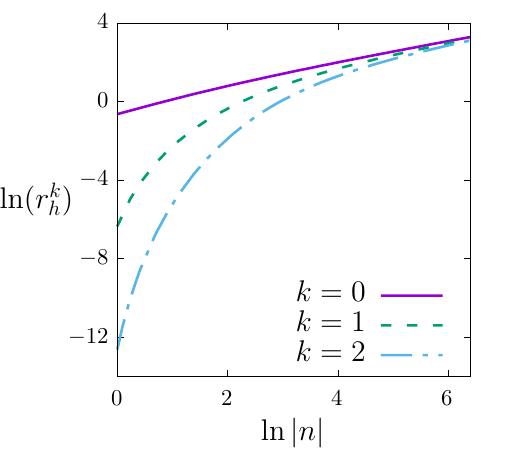}
                \includegraphics[scale=0.50]{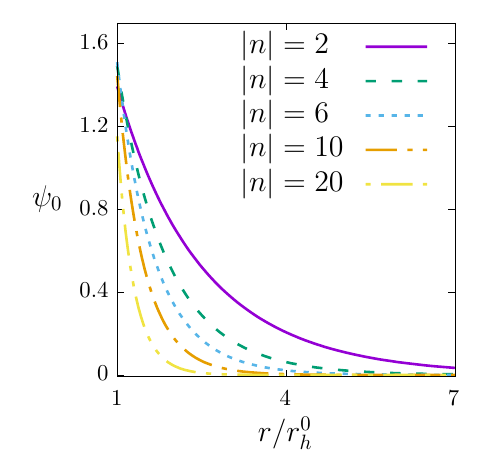}
    
         \caption{\small The RN horizon size $r_h^k(n)$ for which there exists a $k$-node zero mode $\psi_{k}(r)$ (left) and 
         the profile of the fundamental zero mode $\psi_0(r)$ (right). 
         }
    \label{Figw}
\end{figure} 
The WS equations \eqref{P2} assume the form 
\be           \label{eqWS}
\nabla^\mu F_{\nu\mu}=4\pi J_\nu\,, ~~~~~
\nabla^\mu Z_{\nu\mu}+\frac{\phi^2}{2}\,Z_\nu=-4\pi\,\frac{g}{g^\prime}\, J_\nu\,, \nn \\
D^\mu w_{\mu\nu}+i\left( g^2\psi_{\nu\sigma}+gg^\prime F_{\nu\sigma}-g^2 Z_{\nu\sigma}  \right)w^\sigma
=\frac{g^2\phi^2}{2}\, w_\nu\, ,\nn \\
\nabla^\mu\nabla_\mu \phi=\frac14\left( g^2\,w_\mu \bar{w}^\mu+Z_\mu Z^\mu\right)\phi+\frac{\beta}{4}(\phi^2-1)\phi\,. ~~~~
\ee
Here the gauge covariant derivative can be written in the form 
$D_\mu=\nabla_\mu+i( e A_\mu- g^2 Z_\mu)$ and 
the current is 
\be          \label{cur33}
4\pi J_\nu= e\,\nabla^\sigma \psi_{\sigma\nu} + e\,\Im (\bar{w}^\sigma w_{\sigma\nu}).
\ee

It should be emphasized that $F_{\mu\nu}$ has the potential, $A_\mu$, while 
 $\FF_{\mu\nu}$ given by  \eqref{Nambu} 
and used in the main text to define ${\mathcal J}_\nu$, $\tilde{{\mathcal J}}_\nu$ 
 is not always closed,
\be               \label{fcal}
\FF_{\mu\nu}=\partial_\mu A_\nu-\partial_\nu A_\mu+ e\,\psi_{\mu\nu}.
\ee
This is exactly the same relation as in \eqref{FF} because the tensor $\psi_{\mu\nu}$ defined in \eqref{Hooft} 
reduces in the unitary gauge precisely to $\psi_{\mu\nu}$  defined in \eqref{psi}.

Taking the divergence of  \eqref{fcal} and comparing with \eqref{cur33} yields 
\be          \label{cu1r}
4\pi {\mathcal J}_\nu=\nabla^\mu \FF_{\nu\mu}=\nabla^\mu F_{\nu\mu}+e\nabla^\mu\psi_{\nu\mu}\nn \\
=4\pi J_\nu-e\nabla^\mu\psi_{\mu\nu}\
= e\,\Im (\bar{w}^\sigma w_{\sigma\nu}).
\ee
hence 
\be          \label{cur22}
J_\nu=\J_\nu+\frac{e}{4\pi}\,\nabla^\sigma \psi_{\sigma\nu}.
\ee

Eqs.\eqref{eqWS} admit the solution 
\be        \label{back}
A_\mu dx^\mu=-\frac{n}{2e}\,\cos\vartheta \, d\varphi,~~Z_\mu=w_\mu=0,~~\phi=1,~~~~~~
\ee
assuming the background geometry to be RN. Expanding Eqs.\eqref{eqWS} around this solution 
yields in the first order of perturbation theory 
\be            \label{Proc}
D^\mu w_{\mu\nu}+ieF_{\nu\sigma}w^\sigma
=\frac{g^2}{2}\, w_\nu\, ,\
\ee
with $D_\mu=\nabla_\mu+ieA_\mu$ where $A_\mu$ is the same as in \eqref{back}
(the fields $Z_\mu,\phi$ also do not change in the first order). 
This equation admits a  solution of the form
\be             \label{ww}
&&w_\mu dx^\mu=e^{i\omega t}\sum_{{\rm m}\in[-j,j]} c_{\rm m} w_{\rm m}(r,\vartheta,\varphi),~~~\\
&&{w}_{\rm m}=  \psi(r)\, 
(\sin\vartheta)^{j}\left(\tan\frac{\vartheta}{2}\right)^{\rm m} e^{-i{\rm m}\varphi}\,
(d\vartheta-i\sin\vartheta d\varphi),~~~~~~~\nn
\ee
if $n>1$, while for $n<-1$ one has to replace $w_{\rm m}\to  \bar{w}_{\rm m}$. 
The value $|n|=1$ is excluded since $w_{\rm m}$ is then unbounded at the symmetry axis. 
One has $j=|n|/2-1$. 
The coefficients $c_{\rm m}$ are arbitrary and can be chosen to be real.  The
function $\psi(r)$ fulfills the equation
\be             \label{RNss}
\left(-\frac{d^2}{dr_\star^2}+N(r)\left[\mw^2-\frac{|n|}{2r^2}\right]\right)\psi(r)=\omega^2 \psi(r),~~~~~
\ee
with  $dr_\star=dr/N(r)$ where $N(r)=1-2M/r+Q^2/r^2$ and $Q^2=(\kappa n^2)/(8e^2)$.
 This equation admits normalizable solutions with $\omega^2<0$
(negative modes) if the horizon radius $r_h=M+\sqrt{M^2-Q^2}$ is small. Therefore, small RN black holes are 
unstable with respect to  the formation of  the charged W-condensate, whereas large black holes are stable. 

Varying $r_h$, one can detect critical values $r^k_h(n)$ for which instabilities just settle in as zero modes, that is 
normalizable solutions with $\omega=0$. These solutions, $\psi_k(r)$, are characterized by 
$k=0,1,2,\ldots $ nodes and they exist 
only for discrete values of the event horizon radius $r_{h}=r^k_h(n)$ shown in Fig.\ref{Figw}. 
The functions $\psi_k(r)$ 
are finite at the horizon and decay exponentially fast 
at large $r$; see Fig.\ref{Figw}. Since $\omega=0$, the solution \eqref{ww}  with  $\psi(r)=\psi_k(r)$ describes static deformations
of the RN black hole, the value $k=0$ corresponding to the fundamental mode and $k>0$ being the radial excitations. 

Setting in  \eqref{ww} $\omega=0$, the solution 
can be transformed to the form expressed by Eq.\eqref{w} of the main text, while \eqref{RNss} reduces to Eq.\eqref{RNs} of the main text. 
This solution 
is defined up to an overall sign which 
can be flipped by the gauge transformation generated by ${\rm \U}=-\tau_3$,
whose effect is $\T_1\to-\T_1$ and $\T_2\to-\T_2$, hence $w_\mu\to -w_\mu$  while the RN background does not change. 
It follows that there are two options: either $w_\mu$ does not change under $\varphi\to\varphi+2\pi$ or it flips sign. 
Therefore, since 
$w_{\rm m}\sim\exp(-i{\rm m}\varphi)$, the azimuthal number ${\rm m}$ can assume either only integer or only 
half-integer values. 
For example, if $j=1$ then the two options are  either ${\rm m}=\pm 1/2$ or ${\rm m}=0,\pm 1$. 

There is always an option containing the value ${\rm m}=0$ and it is  possible to choose $c_{\rm m}=\delta_{{\rm m}0}$
since coefficients $c_{\rm m}$ are arbitrary within the linear perturbation theory. 
However, they are no longer arbitrary when the higher order corrections are taken into account, 
although the possibility to choose $c_{\rm m}=const.\times \delta_{{\rm m}0}$ always exists.

One has $\Im (\bar{w}^\sigma w_{\sigma\nu})=0$ for the solution \eqref{ww} hence 
$
4\pi J_\mu= e\nabla^\sigma \psi_{\sigma\mu}\, 
$
 is tangential to the sphere, with 
non-zero components 
\be             \label{J}
J_\vartheta= \frac{e }{8\pi\sin\vartheta}\,\partial_\varphi |w_\mu|^2,~~
J_\varphi= -\frac{e \sin\vartheta }{8\pi}\,\partial_\vartheta |w_\mu|^2,~~~~~~~
\ee
where 
\be        \label{PSI}
|w_\mu|^2&\equiv& w_\mu \bar{w}^\mu=\frac{2}{r^2}\,|\psi(r)|^2\times  \Theta(\vartheta,\varphi), \\
\Theta(\vartheta,\varphi)&=&\left(\sin\vartheta\right)^{2j}\sum_{\rm k,m}c_{\rm k}c_{\rm m}\left(\tan\frac{\vartheta}{2}\right)^{\rm k+m}
e^{i\rm(k-m)\varphi}\,.~~~~\nn
\ee
It follows from \eqref{J} that $J^\mu$ is orthogonal to the gradient of  $|w_\mu|^2$, 
which is in turn orthogonal to the level lines of $|w_\mu|^2$. Therefore,  the latter are parallel to  $J^\mu$.

The current is quadratic in $w_\mu$, $c_{\rm m}$ and sources second order correction for the electromagnetic field, $f_{\mu\nu}$,
second order $Z_\mu$ and second order $\delta\phi=\phi-1$. Expanding \eqref{eqWS} up to second order terms yields 
\begin{subequations}               \label{eqWS2}
\begin{align}
\nabla^\mu f_{\nu\mu}&= e\, \nabla^\sigma \psi_{\sigma\nu}\,, \label{e1} \\ ~~~~~
\nabla^\mu Z_{\nu\mu}+\frac{1}{2}\,Z_\nu&= -g^2\nabla^\sigma \psi_{\sigma\nu}\,, \label{e2}  \\
\nabla^\mu\nabla_\mu \delta\phi-\frac{\beta}{2}\,\delta\phi&= \frac{g^2}{4}\,|w_\mu|^2 \,. \label{e3}
\end{align}
\end{subequations}
Solving these equations and injecting to \eqref{LEW1},  yields the result containing terms up to fourth order, 
\be           \label{LEW2}
-{\cal L}_{\rm EW}&=&\frac{1}{4}(g F_{\mu\nu}+g^\prime Z_{\mu\nu})^2 
+\frac{1}{4}(g^\prime F_{\mu\nu}-g Z_{\mu\nu}+ g\,\psi_{\mu\nu})^2 \nn \\
&&+ \frac{1}{4}\,|w_{\mu\nu}|^2 
+ \frac{g^2}{4}\,|w_\mu|^2(1+2\delta\phi)+\frac14\,(Z_\mu)^2 \nn \\
&&+(\partial_\mu\delta \phi)^2+\frac{\beta}{2}\,\delta\phi^2. 
\ee
Here $F_{\mu\nu}=\overset{0}{F}_{\mu\nu}+f_{\mu\nu}$ with $\overset{0}{F}_{\mu\nu}$ being the zeroth-order field 
corresponding to \eqref{back}. Since the fields are static and purely magnetic, the energy density is ${\cal E}=-{\cal L}_{\rm EW}$. 

The procedure now is to solve Eqs.\eqref{eqWS2}, compute the energy $E=\int {\cal E}\sqrt{\rm -g}\, d^3x$, and then minimize it 
with respect to the coefficients $c_{\rm m}$ by keeping fixed the norm $\int w_\mu \bar{w}^\mu \,\sqrt{-\rm g}\, d^3x$. 
Eqs.\eqref{eqWS2} can be solved with Green's functions, as was done in \cite{Ridgway:1995ke} where a simplified theory 
without $Z$, $\Phi$ was considered. However, we postpone this  for future work since our main goal 
at present is to study axially-symmetric solutions. To get some  idea of more general solutions, we use  an approximate procedure,
which nevertheless seems to capture   the essential points. 
The key observation is that the energy density always contains the fourth order term $(\psi_{\mu\nu})^2\sim |w_\mu|^4$. 
Therefore, we minimize 
\be
\langle |w_\mu|^4\rangle\equiv \int |w_\mu|^4 \,\sqrt{-\rm g}\, d^3x\,,
\ee
under the condition that the norm should be fixed, 
\be
\langle |w_\mu|^2\rangle\equiv \int |w_\mu|^2 \,\sqrt{-\rm g}\, d^3x=const.
\ee
One can also argue as follows. 
Eq. \eqref{e1} is fulfilled by 
 \be             \label{approx}
 f_{\mu\nu}=- e\, \psi_{\mu\nu}. 
 \ee
 This solution is not totally satisfactory since the form $\psi_{\mu\nu}$ is not closed,
 but we ignore this. 
 Then the non-zero components are 
\be             \label{vort}
f_{\vartheta\varphi}=-f_{\varphi\vartheta}=e\,\frac{n}{|n|}\,|\psi(r)|^2 \Theta(\vartheta,\varphi)\sin\vartheta,
\ee
which corresponds to a radial magnetic field of the same direction as the background field. 
This confirms  the EW anti-Lenz law: the magnetic field induced  by the condensate enhances the external  field
and is maximal where the condensate is maximal 
\cite{Ambjorn:1988tm,Ambjorn:1989sz,Chernodub:2012fi,Chernodub:2022ywg}.

Injecting \eqref{approx} to \eqref{LEW2}, integrating by parts, using Eqs.\eqref{eqWS2}, and  omitting the zeroth order order term  yields 
\be            \label{EN}
{\cal E}= \frac{g^4}{4} (\psi_{\mu\nu})^2+ \frac{1}{4}\, |w_{\mu\nu}|^2+ \frac{g^2}{4}\, |w_\mu|^2 \nn \\
- \frac{g^2}{4}\, \psi^{\mu\nu}Z_{\mu\nu}+ \frac{g^2}{4}\, |w_\mu|^2\delta\phi, 
\ee
where 
\be
(\psi_{\mu\nu})^2=\frac{1}{2}|w_\mu|^4,~~|w_{\mu\nu}|^2=2N(r)\left|\frac{\psi^\prime(r)}{\psi(r))}\right|^2\,|w_\mu|^2.~~~~~~
\ee
We do not know what $Z_{\mu\nu}$ and $\delta\phi$ are, but if we simply ignore them  and set to zero the second line in \eqref{EN},
then the integration gives 
\be            \label{ENN}
E=const.\times \langle |w_\mu|^4\rangle+const.\times \langle |w_\mu|^2\rangle. 
\ee
The same result is obtained if we  ignore only $Z_{\mu\nu}$ but use the approximate solution for $\delta\phi$ obtained by neglecting the 
derivatives in \eqref{e3}, 
$
\delta\phi= -g^2/(2\beta)\,|w_\mu|^2. 
$
Of course, the energy obtained by genuinely  solving the equations \eqref{eqWS2} should have a more complicated structure. 
The justification for the heuristic prescription \eqref{ENN} 
is that it is simple and gives what is expected -- the ``platonic'' distribution of vortices on the horizon. 
Therefore, it seems that the prescription \eqref{ENN} gives correct results, even though it relies on approximations. 

Using \eqref{PSI}, the second term in \eqref{ENN} reduces to $const.\times E_2$ with 
\be              \label{E2}
E_2=\sum_{{\rm m}\in[-j,j]} A_{j,\rm m}\, c_{\rm m}^2\,,
\ee
where 
\be
A_{j,\rm m}&=& \int_0^\pi  (\sin\vartheta)^{2j+1} \left(\tan\frac{\vartheta}{2}\right)^{\rm 2m} d\vartheta \nn \\
&=&2^{2j+1}\frac{\Gamma(j+1+{\rm m})\Gamma(j+1-{\rm m})}{\Gamma(2j+2)}\,.
\ee
Similarly, 
 after some algebra, 
the first term in \eqref{ENN} reduces to  $const.\times E_4$ with 
\be
E_4= \sum_{\rm k,m,l}A_{2j,\rm k+l}\, c_{\rm m}\, c_{\rm k}\, c_{\rm l}\, c_{\rm k+l-m}\,.
\ee
It is assumed  here that $c_{\rm m}=0$ if $|m|>j$. 
Therefore, omitting constant factors, the problem reduces to minimization  
with respect to $c_{\rm m}$ of
\be         \label{Emin} 
E=E_4+\mu\,(E_2-1),
\ee
 where $\mu$ is  the Lagrange multiplier. 

Assuming for example that  $n=10$ hence $j=4$ and $m=-4,-3,\ldots 4$,  so that there are 9 coefficients $c_{\rm m}$ to find. 
Their values corresponding to the absolute minimum of $E$ determine the function $|w_\mu|^2$ in \eqref{PSI},
whose value at the horizon is shown in Fig.\eqref{Fig0} (left panel) of the main text. The figure shows the contour lines  which, as explained above, 
coincide with the flow lines of the current $J^\mu$.

As seen in Fig.1, flow lines form loops (blue online). A loop of current generates a magnetic field in the orthogonal  
direction, hence the loops in Fig.1  encercle (pieces of) radially directed vortices approximately described by \eqref{vort}. 
 The special feature of these EW vortices is that their magnetic field is enhanced by the condensate: it 
 vanishes at the 
vortex center where the condensate is zero and maximal where the condensate is maximal 
\cite{Chernodub:2012fi,Chernodub:2022ywg}. In other words, these vortices resemble hollow tubes: everything is concentrated at the 
tube surface and nothing is inside.

There are 4 vortices on the visible side of the surface in Fig.1 
and the same number on the other side, so that altogether there are $8=n-2$ vortices homogeneously distributed over the horizon. 
The total magnetic flux is $2\pi\times n$, and the difference $|n|-2$ arises because for $n=2$ the black hole is spherically symmetric \cite{Bai:2020ezy}
hence there are no vortices at all. We checked that minimization of \eqref{Emin} gives one vortex for $n=3$, two vortices for $n=4$, etc. 
These solutions correspond to the global energy minimum. 

The axially symmetric configuration with
 $c_{\rm m}\propto \delta_{\rm m 0}$ is also a stationary point of the energy 
\eqref{Emin}, which can be viewed  as a system in which  all elementary  vortices       
merge into two oppositely directed collective multi-vortices   
 generated by two oppositely-directed azimuthal currents. 

The Hessian matrix has negative eigenvalues in this case, 
therefore such  solutions are unstable and decay  into non-axially symmetric  ones. 
 The exceptional case is $|n|=2$ when only ${\rm m=0}$ is possible;
 the corresponding solution is stable in the flat space limit  \cite{GVI}, while  its gravitating version 
 should very likely to be stable too.
 Another excepltional case is $|n|=4$ when 
the values of $c_{0},c_{\pm 1}$ corresponding to the global energy minimum can be transformed    to
 $c_{\rm m}\propto  \delta_{\rm m 0}$ by a global rotation. Therefore, the axially symmetric configuration 
 is already in the global energy minimum. 
 
 %



\end{document}